\documentclass[acmsmall]{acmart}

\acmJournal{JRC}
\acmVolume{2}
\acmNumber{4}
\acmArticle{16}
\acmMonth{10}
\acmYear{2025}
\acmDOI{10.1145/3748516}
\usepackage[utf8]{inputenc}
\usepackage{booktabs}
\usepackage{array,makecell}
\usepackage{tabularx}
\def\tbl#1#2{\caption{#1}{#2}}
\def\tabnote#1{\par\vspace{2pt}\small #1}
\def\mdash{--}

\makeatletter
\AtBeginDocument{\@ifpackageloaded{natbib}{\ifNAT@numbers\if@filesw\immediate\write\@auxout{\string\global\string\NAT@numberstrue}\fi\fi}{}}
\makeatother

\begin{document}

% \ShortTitle{Constraining Participation: Affordances of Feedback Features in Interfaces to LLMs}
\title{Constraining Participation: Affordances of Feedback Features in Interfaces to Large Language Models}

\author{Ned Cooper}
\orcid{0000-0003-1834-279X}
\email{edward.cooper@anu.edu.au}
\affiliation{\institution{Australian National University}
\city{Canberra}
\state{Australian Capital Territory}
\country{Australia}}

\author{Alexandra Zafiroglu}
\orcid{0000-0003-0918-3977}
\email{alex.zafiroglu@anu.edu.au}
\affiliation{\institution{Australian National University}
\city{Canberra}
\state{Australian Capital Territory}
\country{Australia}}

\begin{abstract}
Large language models (LLMs) are now accessible to anyone with a computer, a web browser, and an internet connection via browser-based interfaces, shifting the dynamics of participation in AI development. This article examines how interactive feedback features in ChatGPT's interface afford user participation in LLM iteration. Drawing on a survey of early ChatGPT users and applying the mechanisms and conditions framework of affordances, we analyse how these features shape user input. Our analysis indicates that these features \emph{encourage} simple, frequent, and performance-focused feedback while \emph{discouraging} collective input and discussions among users. Drawing on participatory design literature, we argue such constraints, if replicated across broader user bases, risk reinforcing power imbalances between users, the public, and companies developing LLMs. Our analysis contributes to the growing literature on participatory AI by critically examining the limitations of existing feedback processes and proposing directions for redesign. Rather than focusing solely on aligning model outputs with specific user preferences, we advocate for creating infrastructure that supports sustained dialogue about the purpose and applications of LLMs. This approach requires attention to the ongoing work of ``infrastructuring''---creating and sustaining the social, technical, and institutional structures necessary to address matters of concern to stakeholders impacted by LLM development and deployment.
\end{abstract}




\authorsaddresses{Authors' Contact Information: Ned Cooper, Australian National University, Canberra, Australian Capital Territory, Australia; e-mail: edward.cooper@anu.edu.au; Alexandra Zafiroglu, Australian National University, Canberra, Australian Capital Territory, Australia; e-mail: alex.zafiroglu@anu.edu.au.}

\maketitle
\renewcommand{\shortauthors}{N. Cooper and A. Zafiroglu}

\section{Introduction}
The deployment of ChatGPT by OpenAI in 2022 marked a significant shift in how \textbf{large language models} (\textbf{LLMs}) are made available to the public. ChatGPT's browser-based interface facilitates broad access to interactions with LLMs through text prompts, expanding user participation in the iterative development of LLMs. Previously, companies such as OpenAI and Google took a more cautious approach. OpenAI used \textbf{application programming interfaces} (\textbf{APIs}) and an iterative approach to deploy the GPT-2 and GPT-3 family of models \cite{Solaiman2019-lt,Brundage2022-ey}, limiting access to developers and researchers. Similarly, Google announced versions of the \textbf{Language Model for Dialogue Applications} (\textbf{LaMDA}) in 2021 and 2022 but did not make it publicly available through an interface or model release \cite{Thoppilan2022-bc}. Following ChatGPT's deployment, numerous other LLMs were made accessible via browser-based interfaces (e.g., Bard/Gemini and Claude.ai) or open releases (e.g., LLaMA).

\looseness-1 This shift in the approach to deployment has precipitated a change in the relationship between companies developing LLMs and their users---we are not only users of LLMs but also participants in their iteration. For many years, users and non-users have been represented in datasets used to train models, constituting a form of passive and involuntary participation \cite{Sloane2022-vh}. However, with recent deployments of public-facing interfaces, users now have a more active and voluntary avenue for participation by submitting feedback on prompt completions through interactive features in these interfaces.

This article focuses on the affordances of interactive feedback features in public-facing interfaces to LLMs, using ChatGPT as a case study. Feedback received through these features plays a role in the iteration of LLMs by aligning model outputs with specific preferences. This ``bottom-up'' approach to alignment elicits preferences from broad groups of users \cite{Gabriel2020-zl}. Others have sought to expand the range of perspectives represented in these groups beyond the limited set of human raters involved in instruction-tuning LLMs before deployment \cite{Kirk2024-dn}. However, our analysis focuses on \emph{how} the affordances of the post-deployment feedback system amplify certain perspectives and concerns while marginalising others. Specifically, we address two research questions:
\begin{enumerate}
  \item What feedback features are present in the ChatGPT interface, and how are they used?
  \item How do these feedback features, through specific mechanisms and conditions, afford particular types of user participation in the iteration of LLMs?
\end{enumerate}

To address these questions, we first surveyed ChatGPT users who have submitted feedback. We then drew on the survey insights to analyse feedback features in the ChatGPT interface that allow users to submit feedback, using Davis's \cite{Davis2020-uq}  mechanisms and conditions framework of affordances. Understanding early adopter use of these feedback features offers insights into their initial affordances and potential limitations when scaling to broader populations; these early patterns are likely to shape the norms and standards for participation in LLM iteration. Our analysis indicates that the ChatGPT interface \emph{encourages} simple, frequent, individualised, and unidirectional feedback on performance-related concerns while \emph{discouraging} feedback from collectives or discussions among users, and \emph{refuses} feedback modification over time. Further, technical background and engagement frequency significantly influence participation patterns.

\looseness-1 We discuss implications arising from the constraints on user participation observed in our study for (re)designing feedback for LLMs. First, we explore deliberative approaches that might enable bidirectional communication between users and developers, changes in perspectives over time, and discussion about social or ethical issues among groups of users. Second, we consider the ``infrastructuring'' \cite{Le_Dantec2013-zu} that could support more expansive feedback channels, to enable sustained dialogue with stakeholders affected by LLM deployments about the purposes and applications of these technologies. Moving beyond simple preference elicitation toward more expansive feedback channels appears increasingly important as LLMs become ubiquitous and unquestionable infrastructure---systems embedded in other structures, transparent in use, and only visible upon breakdown \cite{Star1996-gz}.

\section{Background}
In this section, we explore concepts related to participatory approaches, feedback features, and the challenges of scale in the context of LLMs. First, we provide an overview of the alignment problem and the ``participatory turn'' in AI \cite{Delgado2023-tm}. Next, we examine feedback features used in recent deployments of LLMs through browser-based interfaces. Then, we explore the implications of scale for the ``sameness'' of those features across diverse contexts \cite{Young2024-ak}. We conclude the section by proposing the study of these concepts through the mechanisms and conditions framework of affordances.

\subsection{Alignment and Participatory Approaches}\label{participatoryAI}
In recent years, technology companies have explored various forms of user participation to address the alignment problem---ensuring that AI systems behave in ways that align with human values \cite{Russell2019-bu}. Two fundamental questions pertinent to the alignment problem are: what values should AI systems align with, and who should decide those values? \cite{Gabriel2020-zl} One approach to addressing both questions is bottom-up---eliciting preferences from a broadly defined public and inferring from those preferences the values to align with \cite{Gabriel2020-zl}.

Interest in participatory approaches for aligning LLMs is part of a broader participatory turn in AI research, design, and development \cite{Delgado2023-tm}. Participatory AI projects involve end users and those impacted by AI systems in the design and development of those systems \cite{Cooper2024-wt}, drawing on long-standing traditions in other fields, such as participatory design \cite{Muller1993-ox,Sundblad2011-wc} and participatory planning \cite{Arnstein1969-ku,Kelty2020-tt}. Participatory AI projects now feature in articles, panels and workshops at prominent computer science conferences \cite{Kulynych2020-cp,Zytko2022-wo} and in recent policy frameworks \cite{Farthing2021-iv,US_Office_of_Science_and_Technology_Policy2022-pl,Tabassi2023-dt}.

\looseness-1 The participatory turn is one response to calls for moves away from formal models of fairness \cite{Green2022Escaping} toward situated collaborations with communities affected by AI systems \cite{Sloane2022Make}. These calls reflect a growing recognition that algorithmic fairness is not merely a technical problem amenable to formal solutions, but a socio-political challenge requiring attention to structural inequities \cite{Hoffmann2019Where,Wong2020-fc}. Indeed, technical approaches alone may obscure or reinforce structural inequities, leading some to advocate for alternatives to algorithmic fairness such as ``algorithmic reparation'' \cite{Davis2021Algorithmic}.

Participatory approaches promise more ``democratic'' AI systems \cite{Kulynych2020-cp} and more ``pluralistic'' alignment \cite{Kirk2024-dn}. However, there is a broad spectrum of participatory practices in AI design, from informing or consulting participants to equitable partnerships with participants \cite{Birhane2022-fk,Corbett2023-de,Delgado2023-tm}. For the development of large-scale AI systems, participation has often taken passive and involuntary forms through the representation of groups of individuals in datasets used to train models \cite{Denton2020-ry,Sloane2022-vh}. While feedback features in interfaces to LLMs provide a more active and voluntary form of user participation, their emergence as the primary mode for publics to influence development processes has raised a question about the conceptual leap from structured preference elicitation to genuine participatory engagement \cite{Feffer2023-jx}.

Additionally, there is a significant gap between the scale of participatory methods in computing research and the scale at which recent AI systems are trained and deployed \cite{Diaz2023-lr}. Surveys of participatory approaches in computing research \cite{Cooper2022-gb} and participatory AI projects \cite{Delgado2023-tm} indicate that engagements have primarily been localised, focusing on one or a few groups or communities with shared characteristics. In contrast, recent deployments of LLMs through browser-based interfaces are accessible to anyone with access to a computer, a web browser, and an internet connection.

\subsection{Interactive Feedback Features in Interfaces to LLMs}
Feedback has a rich history rooted in cybernetics, which introduced the idea of feedback loops, where the output of a system is fed back as input, and queried the implications of feedback loops for modelling dynamical systems over time \cite{Wiener1948-rg}. The concept of feedback loops influenced subsequent AI research directions \cite{Minsky1961-sg} and remains fundamental to AI systems that learn over time. Concurrently, ``feedback'' has also become a general term referring to information provided by an observer (e.g., users or other stakeholders) assessing a system's performance.

\textbf{Reinforcement Learning from Human Feedback} (\textbf{RLHF}) combines these feedback types in modern AI development. RLHF begins with evaluative feedback, when a model generates two responses to a given prompt and human raters determine which response is better than the other according to some criteria. These human assessments are then used to create a cybernetic feedback loop: an RL algorithm uses the assessments to predict how a human would rate other responses, and an LLM is fine-tuned on the resulting reward model \cite{Ziegler2019-me,Ouyang2022-rc,Bai2022-qx}. While central to aligning pre-trained LLMs with specific preferences \cite{Lambert2023-pr}, integrating human feedback into the learning process increases training time and labour costs \cite{Maslej2024-xr}. RLHF requires large, diverse groups of human raters to assess prompt completions before fine-tuning the model \cite{Prabhakaran2021-pe}.

The growing importance of RLHF has increased demand for human feedback data \cite{Lambert2023-pr} and research attention on practices for creating and curating diverse, representative, and quality human feedback data \cite{Feffer2023-am,Kirk2024-dn,Prabhakaran2021-pe}. However, less attention has been paid to human feedback collected through interfaces to LLMs, in which users evaluate prompt completions of deployed systems. This post-deployment feedback may be used to improve future model iterations. For example, OpenAI used feedback on non-factual responses obtained through the feedback features in ChatGPT's interface, with additional labelled comparison data, to train reward models that reduced GPT-4's tendency to ``hallucinate'' compared to GPT-3.5 \cite{OpenAI2023-iq}.

There is growing interest in crowdsourcing human feedback datasets to align LLMs with user preferences \cite[e.g.,][]{LeCun2023-yq}. In the blog post announcing the deployment of ChatGPT, OpenAI emphasised their interest in collecting user feedback to uncover potential risks and harms:

\begin{quote}
``We are ... interested in feedback regarding harmful outputs that could occur in real-world, non-adversarial conditions, as well as feedback that helps us uncover and understand novel risks and possible mitigations.'' \cite{OpenAI2022-oj}
\end{quote}

OpenAI launched a feedback competition alongside the ChatGPT interface to encourage user participation (running through December 2022). Users were encouraged to submit free-form feedback via a Google form that could help OpenAI identify potential real-world risks and harms, with winners receiving \$500 in OpenAI API credits.

\begin{figure}[!t]
\vspace*{2pt}
\centering
\includegraphics[width=0.5\linewidth]{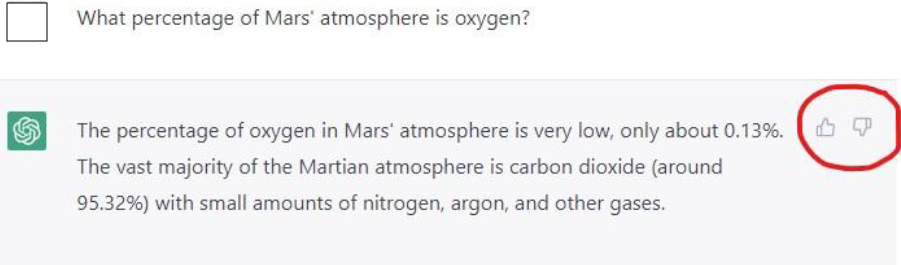}
\caption{Thumbs-up or down feedback - ChatGPT (May 2023).}
\Description{An image of the ChatGPT interface showing a prompt completion. To the right of ChatGPT's response, there are two buttons side by side: a thumbs-up icon on the left and a thumbs-down icon on the right. This represents the simple binary feedback mechanism offered to users after each prompt completion.}
\label{fig:thumbs23}
\end{figure}

\begin{figure}[!t]
\centering
\includegraphics[width=0.5\linewidth]{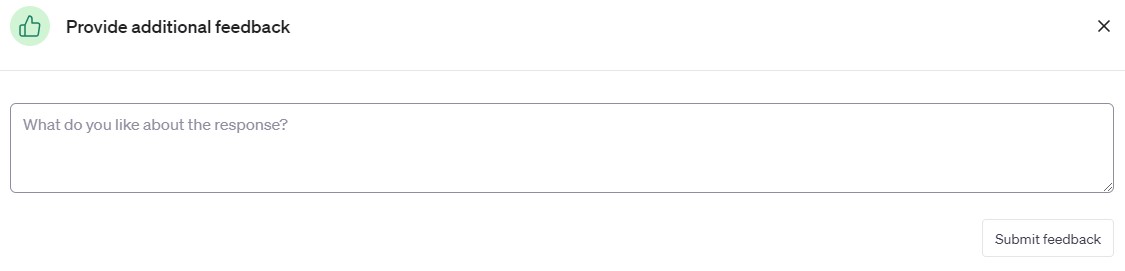}
\caption{Open-text window after thumbs-up - ChatGPT (May 2023).}
\Description{An image of a pop-up window in the ChatGPT interface that appears after a user gives a thumbs-up rating. The window is titled ``Provide additional feedback'' and contains a large text box for users to input detailed feedback, in which the following text appears: ``What do you like about the response?''. Below the text box to the right is a ``Submit feedback'' button.}
\label{fig:open_pos23}
\end{figure}

\begin{figure}[!t]
\vspace*{2pt}
\centering
\includegraphics[width=0.5\linewidth]{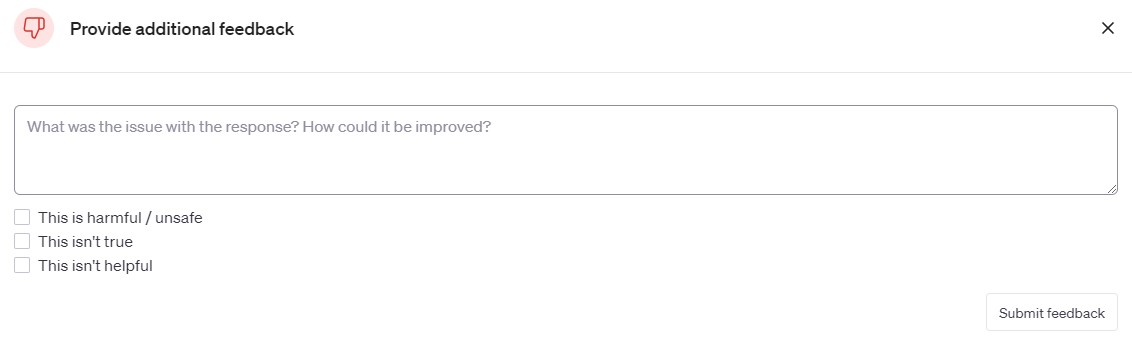}
\caption{Open-text window after thumbs-down - ChatGPT (May 2023).}
\Description{An image of a pop-up window in the ChatGPT interface that appears after a user gives a thumbs-down rating. The window is titled ``Provide additional feedback'' and contains a large text box for users to input detailed feedback, in which the following text appears: ``What was the issue with the response? How could it be improved?''. Below the text box are three checkboxes labelled ``This is harmful/unsafe'', ``This isn't true'', and ``This isn't helpful''. At the bottom right of the window is a ``Submit feedback'' button.}
\label{fig:open_neg23}
\end{figure}

\begin{figure}[!t]
\centering
\includegraphics[width=0.5\linewidth]{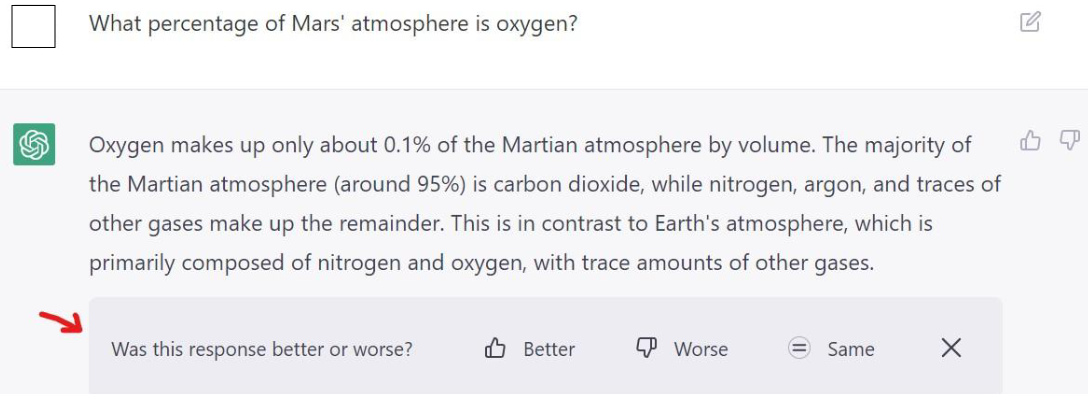}
\caption{Feedback after regenerating a response - ChatGPT (May 2023).}
\Description{A screenshot of the ChatGPT interface showing a regenerated response. Between the responses is a set of three buttons arranged horizontally, allowing the user to rate whether the new response is ``Better'', ``Worse'', or the ``Same'' compared to the original response.}
\label{fig:regen23}
\end{figure}

The ChatGPT interface was also equipped with several interactive features, including: thumbs-up/thumbs-down ratings (see Figure~\ref{fig:thumbs23}) open-text feedback (see Figure~\ref{fig:open_pos23} and Figure~\ref{fig:open_neg23}), and comparative feedback after a user regenerates a response (see Figure~\ref{fig:regen23}). Checkboxes in the open-text feedback windows allow users to structure their feedback against the prominent ``helpful, honest, and harmless'' framework for AI alignment \cite{Askell2021-yq}. Similar features appear in other interfaces to LLMs, such as Google's Bard/Gemini and Anthropic's Claude.ai. The deployment of LLMs through these interfaces has made these systems accessible to a vast and diverse user base, presenting new challenges related to scale.

\subsubsection{Scale and Affordances}
The standardisation of feedback features across interfaces to different systems exemplifies the proceduralisation and replication that characterise other digital systems deployed at large scale \cite{Gillespie2020-ib}. LLMs have grown exponentially in size in recent years---from billions of parameters in GPT-2 \cite{Solaiman2019-lt} to (reportedly) trillions in GPT-4 \cite{Patel2023-jc}. But more than simply expanding the size of a system, scaling digital systems across global contexts invariably requires imposing a standardised worldview and organising phenomena that can operate context-independently \cite{Tsing2012-bq,Hanna2020-hv}. This sameness is achieved through massive infrastructural investments, such as data centres, standardised software and policies, and global distribution networks \cite{Young2024-ak}. Feedback features (and the associated infrastructure for managing, storing, and processing feedback data) represent one such infrastructural investment designed to elicit participation at scale. Notably, these feedback features are identical for all users of a given interface to an LLM (e.g., ChatGPT, Bard/Gemini, or Claude.ai) and similar across different interfaces to various LLMs, despite the diversity of global contexts in which users interact with these systems.

\looseness-1 However, this sameness can obscure the power dynamics inherent in the feedback process. Achieving scale in digital contexts is an aggregation of power---engineering the suppression of context to render complex social realities coherent through a computationally legible frame \cite{Tsing2012-bq,Gillespie2020-ib}. As Eubanks \cite{Eubanks2017-pr} demonstrates, the design of scaled technical systems can reinforce existing inequalities and marginalise certain voices. For example, people less likely to engage with these interfaces, such as those with limited digital literacy or access, may be underrepresented in the feedback data used to refine the family of models underlying the interfaces. The standardisation of feedback features across different systems might also lead to the homogenisation of user input, risking the amplification of certain biases or the suppression of diverse perspectives. The ``helpful, honest, and harmless'' framework \cite{Askell2021-yq}, for example, is commonly used to structure feedback for RLHF workflows. However, applying such standardised frameworks risks imposing particular values across diverse contexts \cite{Gabriel2025Matter}. Definitions of ``helpfulness'' can vary culturally \cite{Sambasivan2021-il}, and ``honesty'' privileges certain epistemic standards inherent in a model's training data \cite{Denton2020-ry}, while determining ``harmlessness'' involves complex and contested judgments about potential\break harms \cite{Weidinger2021Ethical, Hoffmann2019Where, Gabriel2025Matter}.

It is thus imperative to critically examine what kinds of perspectives and contexts these feedback features make visible and valuable as forms of participation---both to those who participate and to those who integrate feedback from participants into ML workflows. The concept of affordances provides one useful lens for conducting such an analysis.

\subsubsection{Mechanisms and Conditions Framework of Affordances}
Affordance theory states that an animal perceives its environment not only in terms of the shapes of objects or the space between objects but also in the possibilities of objects for action \cite{Gibson1977-cg}. Animals share the same environment but perceive it differently according to their affordances---their capacity to benefit or experience harm from a particular feature of the environment. Norman \cite{Norman2008-ig} later adapted affordance theory for technology design, arguing that technological objects afford different possibilities for humans and other agents and that those possibilities cannot be defined except by reference to the object's adoption and use \cite{Hildebrandt2015-el}. The affordances of a technological object are thus the link between the object's features and the outcomes of the technological object's adoption and use by subjects \cite{Davis2020-uq}. \citet{Davis2016-wg} describe these links as ``relational processes among users, designers, environments, and things''.

\citet{Davis2020-uq} operationalises the concept of affordances through a framework of mechanisms and conditions. The framework extends affordance theory by reframing the fundamental question away from what technologies afford to ``\emph{how} technologies afford, \emph{for whom}, and \emph{under what circumstances}'' \cite{Davis2020-uq}. Mechanisms describe the link between features of technological objects and the outcomes of their use by subjects, while conditions describe how affordances vary across different subjects and contexts.

The framework proposes six mechanisms that contribute to the operationalisation of affordances: \emph{request}, \emph{demand}, \emph{encourage}, \emph{discourage}, \emph{refuse}, and \emph{allow} \cite{Davis2020-uq}. These mechanisms are intended to serve as a tool for critical analysis rather than as a taxonomy of the features of technological objects, as each mechanism may reveal different aspects of a technical feature. Technological objects \emph{request} and \emph{demand} how user-subjects interact with them. \emph{Requests} refer to bids placed by the artifact initiating some action or set of actions, while \emph{demands} represent more forceful bids that leave users with little room for non-compliance. Technological objects also \emph{encourage} and \emph{discourage}, \emph{refuse} and \emph{allow} actions of user-subjects. These latter mechanisms represent responses from the artifact to activities that users seek to enact. \emph{Allow} functions neutrally, applying to bids placed by both the artifact and user-subjects engaging with it. Importantly, these mechanisms are not mutually exclusive but operate simultaneously---features that \emph{request} certain feedback formats may \emph{encourage} some user behaviour at the same time as \emph{discouraging} others.

\enlargethispage{-7pt}
The mechanisms operate through three interrelated conditions: \emph{perception} (how a subject perceives the artifact based on their skill and technical literacy), \emph{dexterity} (what a subject can do with the artifact), and \emph{cultural and institutional legitimacy} (the social and structural position of the user within institutions of power) \cite{Davis2020-uq}. The interaction between mechanisms and conditions creates variations in how technologies afford actions for different users. For example, without sufficient dexterity or the requisite cultural and institutional legitimacy, what may be \emph{allowed} becomes \emph{discouraged} or \emph{refused} \cite{Davis2023-xk}. The framework thus addresses binarism (assuming an artifact affords some function or not) and universalism (assuming an artifact operates identically for all) in affordance theory \cite{Davis2016-wg}.

If we recognise ML-enabled technologies as material artefacts that reflect and affect users through design decisions, and acknowledge their potential to exacerbate inequality in social settings and workplaces \cite{Davis2023-xk}, we must scrutinise the affordances of such technologies and applications. The mechanisms and conditions framework provides a structured approach to analyse how the design of feedback features in interfaces to LLMs affords certain actions while constraining others, thereby shaping the nature and scope of user input and participation in RLHF workflows. Furthermore, the framework offers a vocabulary for the (re)design of feedback features to address issues we may identify for user input and public participation through socio-technical analysis \cite{Davis2020-uq}.

\section{Method}
We conducted an online survey to gather insights from ChatGPT users who had submitted feedback using the features in the browser-based interface. Through the survey, we aimed at exploring feedback practices and perspectives of ChatGPT users to understand which users submit feedback, how they submit feedback, and why they submit feedback. We then drew on the insights from the survey to analyse the feedback features through the mechanisms and conditions framework of affordances. Below, we detail the process we used to conduct the survey and the analysis, and we then describe the limitations of our approach. In the following section, we present the insights from our analysis.

\subsection{Data Collection}
\subsubsection{Pilot Survey}
We conducted two pilot surveys in March 2023 with 32 respondents from the \textbf{Australian National University} (\textbf{ANU}) to refine our survey instrument's clarity, flow, and length. In late March 2023, the ANU approved our recruitment plan and final survey instrument.

\subsubsection{Survey Instrument}
The survey instrument consisted of 36 substantive questions exploring respondents' frequency and types of ChatGPT usage, experiences submitting feedback through the ChatGPT interface and external channels, reasons for submitting or not submitting feedback, perceptions of the feedback process, the likelihood of providing future feedback, and demographic characteristics (see Appendix A for survey questions). Most questions were multiple-choice, with several open-ended questions included to gather more detailed insights and reflections.

\subsubsection{Recruitment and Sampling}
We administered the public survey on Qualtrics in May 2023 and used a multi-pronged approach to recruit a diverse range of participants. This approach was necessary as the ChatGPT user population was not well-defined at the time of our study. Initially, we posted the survey on our social media accounts (X (formerly known as Twitter), Mastodon, and LinkedIn), forums frequented by active ChatGPT users (GitHub forums, Hugging Face forums, the OpenAI community forum, and Kaggle forums), and on the r/ChatGPT Discord server.

We then expanded recruitment through Prolific to address potential non-response bias and broaden the respondent pool. We applied screening strategies on Prolific to invite suitable respondents to the final survey. Firstly, we used Prolific filters to identify respondents from any country, reporting fluency in English, an approval rate greater than or equal to 95\%, and 10 or more previous submissions. Secondly, we invited filtered respondents to a custom screening survey with two questions asking participants about chatbot usage and feedback submission.\footnote{ChatGPT usage was not an option for filters on Prolific in May 2023.} Finally, we invited respondents approved after the screening survey to the final survey at three separate intervals to cover working hours in regions with different time zones. Respondents to the final survey on Prolific received the equivalent of £9/hr for their time.

\subsubsection{Respondents}
The survey received responses from 526 individuals---after removing incomplete surveys, failed attention checks,\footnote{Only Prolific surveys.} and screened surveys\footnote{Valid IP Address, ``Yes'' to ``Read and agree to Participant Information Sheet'', ``Yes'' to ``Has used ChatGPT''.}---with a median completion time of five minutes and 11 seconds. We recruited 82 respondents through the first recruitment step and 445 through the second. As shown in Table \ref{tab:demographics}, our sample includes perspectives from users with varying levels of technical experience and engagement with AI-related topics (e.g., through news articles, research papers, podcasts). Most respondents did not have formal AI training (73.5\%) or relevant work experience (78.7\%), suggesting our recruitment approach reached beyond technically experienced users. However, respondents were predominantly young, with over three quarters aged 18--34, and the gender distribution was skewed toward men. Geographically, the majority of respondents were from Europe, and the most common languages spoken by respondents were English (UK, US, or AU), Portuguese, and Polish.\footnote{Respondents could select up to three languages they speak daily.}

\begin{table}[!t]%%%1
\centering
\tbl{Respondent Demographics\label{tab:demographics}}{
\renewcommand{\arraystretch}{1.15}
\begin{tabular}{l p{27em}<{\raggedright}}
\hline
\textbf{Characteristic} & \textbf{Distribution} \\
\hline
Age ($n$ = 524) & 18-34 (77.5\%), 35-54 (19.6\%), 55+ (2.3\%) \\
\addlinespace
Gender ($n$ = 521) & Man (63.1\%), Woman (34.9\%), Non-binary/Third gender (1.2\%), Prefer not to say/Prefer to self-describe (0.8\%) \\
\addlinespace
Region ($n$ = 507) & Europe (55.4\%), Africa (15.6\%), UK (10.3\%), Oceania (6.5\%), Other (12.2\%) \\
\addlinespace
AI Experience & Prior use of AI/ML tools (41.0\%, $n$ = 522), Formal education in AI/ML (26.5\%, $n$ = 520), Working in AI/ML field (21.3\%, $n$ = 521) \\
\addlinespace
AI Engagement ($n$ = 523) & Daily/Weekly (63.1\%), Monthly (18.9\%), Rarely/Never (18.0\%) \\
\hline
\end{tabular}}
\end{table}

\subsection{Data Analysis}
Initially, we performed frequency analysis for most questions (see Tables~\ref{tab:chatgpt-usage-frequency}--\ref{tab:feedback-confidence}). For certain variables measured on a five-point Likert scale, we transformed the responses into three categories (e.g., Agree, Neutral, Disagree; or Likely, Neutral, Unlikely) to enhance statistical validity for chi-squared tests (by increasing cell counts and reducing the risk of minimum frequency violations) and improve interpretability. To assess the potential impact of this categorisation on our ordinal Likert scale data we performed sensitivity analyses on key relationships using the original five-point scales. These analyses showed consistent significance patterns in most cases, with similar effect sizes across both approaches.

We then performed chi-squared tests of independence ($\chi^2$) to analyse relationships between responses, using Cramér's V to determine the strength of the associations \cite{Cohen2013-ae}. Before these analyses, we removed rows with missing values for demographics and experience factors and grouped categories with low frequencies for relevant questions (e.g., consolidating age groups and geographic regions). To address the multiple comparisons problem, we applied the Benjamini-Hochberg procedure to control the false discovery rate across all tests simultaneously \cite{Benjamini1995-ir}. We report statistically significant results based on the adjusted $p$-values (adj. $p$-value). All statistical tests were conducted at a 95\% confidence level (df=degrees of freedom; see Appendix C for full results).

For open-ended questions in the survey, we conducted a thematic analysis of responses \cite{Braun2006-cu}, with the first author initially developing codes based on the semantic content of the first 50 responses using tags in the coding software Atlas.ti. Both authors then met to discuss and consolidate the codes, after which the first author coded the remaining responses, followed by a final joint discussion to identify and agree on key themes reflecting the data. Finally, we drew on insights from the survey and our interface analysis to inform the affordances analysis.

\subsection{Limitations}
Our methodology has several important limitations. Firstly, the survey dataset does not statistically represent all ChatGPT users or the subset who submit feedback, limiting the generalisability of our findings \cite{Tufekci2014-nh}. We sampled a subset of users in May 2023, influenced by the authors' networks and the Prolific user population. Additionally, while necessary to engage with users during ChatGPT's early release, sourcing some respondents from dedicated online forums might introduce a bias toward users inclined to discuss experiences publicly, potentially overlapping with those who share feedback through alternative channels. Our multi-pronged recruitment approach was pragmatic given that characteristics of the user base and feedback-submitting population were undefined at the time of study. Consequently, our survey insights are exploratory but useful as a starting point for understanding user practices and perspectives related to feedback features.

\looseness-1 Additionally, our survey relies on self-reported data from respondents. We did not create a primary dataset of feedback instances, as we could not access a dataset of feedback instances submitted through the ChatGPT interface. Instead, we collected a secondary dataset of the type of feedback (e.g., thumbs-up/down, open-text feedback) and the broad categories of feedback (e.g., correcting technical errors/bugs, social and ethical information) provided by respondents. There may be discrepancies between what users reported and their actual feedback behaviour. We note, however, that the survey was conducted primarily to understand the practices and perspectives of ChatGPT users who have submitted feedback rather than to analyse the content of the\break feedback itself.

We also engaged with early adopters of ChatGPT, as our survey was conducted in the first six months after deployment. In April and May 2023, surveys of U.S. adults found that only 16\% \cite{Vogels2023-hn} and 14\% \cite{Berndt2023-ag} had used ChatGPT, respectively. As a result, the respondents' demographics and responses to questions about their experience with AI, ML, and data science may not represent the broader population of ChatGPT users as adoption changes over time. Early adopters tend to have higher education levels, higher socioeconomic status, and be more engaged with developments in emerging technologies than later adopters \cite{Rogers2003-fs}, which may skew the survey results. If ChatGPT usage becomes more mainstream, the characteristics and perspectives of users may change, potentially affecting the types of feedback provided and the motivations for engaging with the feedback features. However, as of March 2024, only 23\% of U.S. adults reported using ChatGPT \cite{McClain2024-oh}, suggesting that the early adopter effect may still be relevant to our findings.

\begin{figure}[!t]
\vspace*{2pt}
\centering
\includegraphics[width=0.5\linewidth]{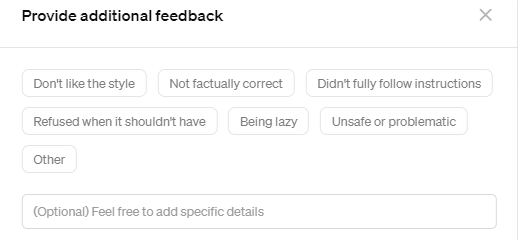}
\vspace*{-3pt}
\caption{Additional feedback categories after selecting thumbs-down - ChatGPT (May 2024).}
\Description{An image of the ChatGPT interface showing a feedback window that appears after a user selects the thumbs-down option. The window is titled ``Provide additional feedback'' and presents multiple categorised feedback options for users to specify why they disliked the response. Categories include ``Don't like the style'', ``Not factually correct'', ``Didn't follow instructions'', ``Refused when it shouldn't have'', ``Being lazy'', ``Unsafe or problematic'', and ``Other''. At the bottom of the window is a single-line text box labelled ``(Optional) Feel free to add specific details'' for users to provide more specific feedback.}
\label{fig:thumbs24}
\end{figure}

\begin{figure}[!t]
\centering
\includegraphics[width=0.5\linewidth]{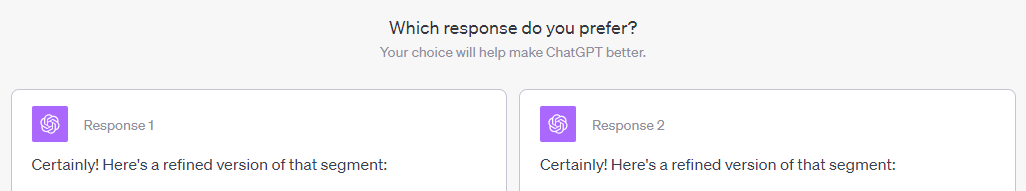}
\caption{Pairwise feedback options - ChatGPT (May 2024, content of responses hidden).}
\Description{An image of the ChatGPT interface displaying a pairwise comparison feature. At the top of the window, the text reads ``Which response do you prefer? Your answer will help make ChatGPT better''. Below, the image shows two text areas side by side containing different generated responses, though the content of the responses is hidden for privacy. This interface design facilitates direct comparison between two alternative responses.}
\label{fig:pairwise24}
\end{figure}

Finally, while survey findings provide useful information about user practices and perspectives concerning feedback, there are limitations to using this data for our affordances analysis. The survey provides a snapshot of user perspectives and experiences at a point in time. The ChatGPT interface and feedback features have and will continue to change, so the affordances of those feedback features and user behaviours may change accordingly. However, as of May 2024, the basic functionality has not changed significantly, though users are prompted with feedback categories where they were previously provided with an open-text box (see Figure~\ref{fig:thumbs24}), and a user may be requested to submit pairwise feedback (see, e.g., Figure~\ref{fig:pairwise24}). We also note that user interviews or observations could provide more in-depth information about how users perceive and interact with the feedback features, though we argue that survey data is sufficient for our high-level analysis through the mechanisms and conditions framework of affordances.

\section{Insights and Analysis}
This section reports insights from the survey regarding ChatGPT usage, feedback practices, and respondents' motivations for submitting feedback. Our analysis indicates how the design of feedback features imposes constraints on end-user participation, operating through various mechanisms and conditions that we explore below. First, we present findings about user interactions with ChatGPT, then examine each affordance mechanism, and finally discuss how these mechanisms operate through different conditions.

Respondents used ChatGPT frequently and in diverse ways. Over 71.6\% ($n$=525) used ChatGPT at least weekly, and the most common use cases were answering specific questions (e.g., factual, technical, or theoretical questions), generating text (e.g., drafting e-mails, writing essays or reports), and academic research assistance (e.g., generating citations, summarising articles).

\subsection{Mechanisms of Affordance}
\subsubsection{Request and Demand}
The ChatGPT interface \emph{requests} feedback formatted in the ``helpful, honest, and harmless'' framework, as evident in the feedback submission options presented to users. This framing guides users toward specific types of evaluative feedback aligned with OpenAI's stated development priorities.

\looseness-1 More fundamentally, the interface \emph{demands} interaction with the system before \emph{allowing} participation through one of the three modes for feedback submission. A user must prompt the dialogue agent before they can submit feedback. This design feature effectively makes system usage a prerequisite for participation in the feedback process, establishing a key mechanism of affordance we observed.

\begin{table}[!t]%%%2
\centering
\tbl{Types of Feedback Submitted ($n$=526)\label{tab:feedback_types_main}}{
\renewcommand{\arraystretch}{1.1}
\begin{tabular}{lrr}
\hline
\textbf{Feedback Type} & \textbf{Submitted} & \textbf{Not Submitted} \\
\hline
Thumbs-up/down feedback & 76.0\% & 24.0\% \\
Open-text feedback & 29.0\% & 71.0\% \\
Feedback after regenerating response & 62.9\% & 37.1\% \\
\hline
\end{tabular}}
\end{table}

\subsubsection{Encourage}
Our survey reveals several patterns in what forms of feedback the interface \emph{encourages}. As shown in Table \ref{tab:feedback_types_main}, a large majority of respondents (84.2\%) submitted at least one form of feedback through the ChatGPT interface. Thumbs-up/down feedback was the most common, followed by feedback after regenerating responses, while open-text feedback was less common.

Furthermore, among those who submitted feedback, respondents submitting thumbs-up/down feedback reported doing so most frequently (28.8\% every/most times), followed by feedback after regenerating a response (27.8\% every/most times), while respondents submitting open-text feedback reported doing so least frequently (15.2\% every/most times) (see Table \ref{tab:feedback_frequency_main}). The observed preference for simpler feedback features may be reinforced by the interface design itself, where the thumbs-up/down feedback icons are visually prominent after each response, while accessing the open-text feedback requires an additional click.

\begin{table}[!t]%%%3
\centering
\tbl{Frequency of Feedback Submission by Type\label{tab:feedback_frequency_main}}{
\renewcommand{\arraystretch}{1.1}
\begin{tabular}{lrrr}
\hline
\textbf{Frequency} & \textbf{Thumbs-up/down} & \textbf{Open-text} & \textbf{After Regenerating} \\
 & ($n$=400) & ($n$=152) & ($n$=331) \\
\hline
Every/Most times & 28.8\% & 15.2\% & 27.8\% \\
About half the time & 24.5\% & 11.2\% & 19.6\% \\
Sometimes & 34.8\% & 50.7\% & 36.3\% \\
Rarely & 12.0\% & 23.0\% & 16.3\% \\
\hline
Total & 100.0\% & 100.0\% & 100.0\% \\
\hline
\end{tabular}}
\end{table}

The interface appears to particularly \emph{encourage} performance-related feedback. Respondents who submitted open-text feedback were primarily motivated by improving the performance and capabilities of the system. We asked this group of respondents to reflect on the most recent time they submitted open-text feedback and report their motivations for doing so. Respondents could select multiple motivations, but those who selected multiple motivations were forced to choose their primary motivation in the following question. Table \ref{tab:feedback_motivations_main} shows the primary motivations in order, taking the weighted frequencies.

\begin{table}[!t]%%%4
\centering
\tbl{Primary Motivations for Submitting Open-text Feedback ($n$=152)\label{tab:feedback_motivations_main}}{
\renewcommand{\arraystretch}{1.1}
\begin{tabular}{lr}
\hline
\textbf{Motivation} & \textbf{Weighted Frequency} \\
\hline
To contribute to AI development & 159 \\
To comment on use-case specific feedback & 94 \\
To report a technical issue or bug & 74 \\
To suggest new features or improvements & 73 \\
To express social or ethical concerns about ChatGPT & 32 \\
Other & 15 \\
\hline
\end{tabular}}
\end{table}

Respondents were also asked what feedback they thought most useful for improving ChatGPT. Based on our thematic analysis, respondents emphasised technical issues (bugs, glitches) and system performance (efficiency, speed). Concerns about accuracy (``hallucinations'' or false information) and the reliability or truthfulness of responses were less common, while social and ethical concerns (harmful, biased, or misleading responses) were mentioned even less frequently.

In summary, while the feedback features in the ChatGPT interface \emph{allow} for open-text feedback, including commentary on social and ethical issues, our respondents predominantly focused on performance-related concerns when submitting feedback. This focus appears to be shaped by the affordances of the feedback system itself rather than by explicitly \emph{discouraging} social or ethical feedback. Despite OpenAI's initial blog post requesting feedback on ``harmful outputs'' and ``novel risks'', our survey respondents appear to have interpreted the purpose of their participation primarily through the lens of improving ChatGPT's performance.

\subsubsection{Discourage and Refuse}
The feedback features in the ChatGPT interface are designed to elicit individual responses, which effectively \emph{discourages} collective feedback or discussions among users. This individualisation extends to the permanence of feedback: the interface \emph{refuses} modification of input once submitted, even if users gain new information or their perspective changes. While users could theoretically input the same prompt a second time and submit feedback, the stochastic nature of LLMs means each response is likely to be unique, making direct comparisons or revisions challenging. This design choice reinforces the individual nature of the feedback process and limits opportunities for evolving, collective input.

Most respondents were ambivalent about whether their reasons for submitting feedback had been addressed---47.4\% neither agreed nor disagreed, 23.7\% somewhat agreed, and 16.4\% somewhat disagreed ($n$=152). Few of those who submitted open-text feedback indicated they had received follow-up communication or updates from OpenAI (11.2\%, $n$=152), though the vast majority of those who had indicated they were satisfied with the response or resolution.

\subsubsection{Allow}
While the ChatGPT interface may constrain direct feedback, it \textit{allows} for alternative channels of expression. Most respondents (80.4\%, $n$=526) did not indicate they shared feedback about ChatGPT through other channels or platforms. However, those who did primarily used X/Twitter (42\%), followed by Facebook (18\%) and Reddit (17\%). Several respondents engaged with dedicated communities, including subreddits like r/ChatGPT and online forums like the OpenAI Developer Forum. Most respondents who indicated they had shared feedback through other channels or platforms (71.9\%, $n$=103) reported they were likely to continue to share feedback through those channels.

\subsection{Conditions}
The mechanisms described above operate through three conditions: perception, dexterity, and cultural and institutional legitimacy. Each condition modifies how mechanisms function for different users, creating variation in how the interface affords participation.

\subsubsection{Perception}
Respondents generally perceived the feedback features in the interface as accessible and user-friendly. As shown in Table \ref{tab:confidence-main}, over three-quarters reported confidence in their ability to evaluate and submit feedback about the performance and/or impact of generative AI systems such as ChatGPT. Of those who submitted at least one of the three feedback modes, almost all found doing so through the features extremely or somewhat easy (95.1\% $n$=442).

\begin{table}[!t]%%%5
\centering
\tbl{Confidence Submitting Feedback ($n$=524)\label{tab:confidence-main}}{
\renewcommand{\arraystretch}{1.1}
\begin{tabular}{lr}
\hline
\textbf{Confidence level} & \textbf{Percentage} \\
\hline
Strongly agree & 21.8\% \\
Somewhat agree & 56.5\% \\
Neither agree nor disagree & 15.1\% \\
Somewhat disagree & 5.7\% \\
Strongly disagree & 1.0\% \\
\hline
\textbf{Total} & \textbf{100.0\%} \\
\hline
\end{tabular}}
\end{table}

However, some users were more confident than others in submitting feedback. Those who frequently engaged with AI-related topics expressed more confidence than would be expected if there was no relationship between these factors and confidence (weak association; see Table \ref{tab:statistical_relationships} for all statistically significant relationships). Conversely, users with less frequent engagement were less confident than would be expected. Younger users (18-34) also showed a weak association with higher confidence in the initial analysis, though this finding was sensitive to the statistical approach used; the association lost significance when tested against the original five-point scale.

The differences in perception impact which mechanisms operate for different user groups. For confident users familiar with AI topics, the interface appears to \emph{encourage} feedback submission through its accessible design. For the 21.8\% of respondents who did not express confidence in evaluating and submitting feedback (those responding ``Neither Agree nor Disagree'' to ``Strongly Disagree''), these same interface features may effectively \emph{discourage} participation despite their availability.

\subsubsection{Dexterity}
Most respondents reported no barriers to accessing ChatGPT or submitting feedback (77.3\%, $n$=519), suggesting that users with varying levels of dexterity can provide input. Among those who encountered barriers, the most common were language and internet connectivity issues. We did not find significant associations between most demographic factors or AI experience and knowledge factors and those who experienced barriers. While those working in AI were more likely to report experiencing barriers than would be expected if there was no relationship between the variables, the association was weak (see Table \ref{tab:statistical_relationships}).

\looseness-1 However, as noted earlier, the requirement to prompt the dialogue agent before submitting feedback effectively \emph{demands} system usage as a prerequisite for participation and \emph{refuses} feedback from non-users. Our sample, therefore, does not include non-user perspectives, and given that most respondents used ChatGPT at least weekly, it also includes limited input from infrequent users. Notably, of those who completed the survey but did not indicate they had submitted feedback through one of the three options, the most common reason was that they were unaware of the feedback options (see Table \ref{tab:reasons_no_feedback_main}). For these users, what designers intended to \emph{allow} is functionally \emph{refused}.

\begin{table}[!t]%%%6
\centering
\tbl{Main Reasons for Not Submitting Feedback ($n$=78)\label{tab:reasons_no_feedback_main}}{
\fontsize{8.7}{9.7}\selectfont
\tabcolsep3pt
\renewcommand{\arraystretch}{1.2}
\begin{tabular}{lrr}
\hline
\textbf{Reason} & \textbf{Count} & \textbf{Percentage} \\
\hline
Unaware of feedback option & 27 & 34.6\% \\
Not enough time or interest in submitting feedback & 21 & 26.9\% \\
Satisfied with performance and have not felt need to submit feedback & 17 & 21.8\% \\
Unsure how to provide feedback & 7 & 9.0\% \\
Other & 5 & 6.4\% \\
Unsatisfied with performance, but do not believe feedback will lead to improvements & 1 & 1.3\% \\
\hline
\textbf{Total} & \textbf{78} & \textbf{100.0\%} \\
\hline
\end{tabular}}
\end{table}

Also, as noted earlier, some respondents found alternative channels and platforms (outside the ChatGPT interface) to express their opinions and participate in discussions about the system. These results suggest that while the interface \emph{discourages} certain forms of participation (e.g., collective feedback or discussions among users), users with sufficient dexterity sought platforms that \emph{allow} forms of engagement the original interface \emph{refuses}.

\subsubsection{Cultural and Institutional Legitimacy}
Feedback features may appeal more to users with certain cultural and institutional backgrounds, which can influence who submits feedback and the types of feedback they provide. Indeed, most of our survey respondents were young and identified as men, consistent with ChatGPT's reported user base \cite{Vogels2023-hn} and studies of contributors to human feedback datasets \cite{Kirk2024-dn}. Generally, feedback types were proportional across demographic groups, though geographic region moderately correlated with open-text feedback submission (see Table \ref{tab:statistical_relationships}).

More significantly, the survey insights in Table \ref{tab:statistical_relationships} suggest that feedback features appeal more to users with frequent engagement or familiarity with AI, ML, and data science. Frequent ChatGPT users, particularly daily users, were more likely to provide thumbs-up/down feedback and those who previously used AI tools were more likely to provide open-text feedback than would be expected if there was no relationship between the variables. We also observed that education in AI correlated with open-text feedback submission, though this relationship fell just short of statistical significance (adj. $p$-value = 0.059, see Table \ref{tab:chi-squared-results}). These patterns suggest that technical background, rather than demographic characteristics, most strongly influenced how our respondents engaged with feedback features.

These insights suggest how cultural and institutional legitimacy conditions shape the operation of affordance mechanisms. While the interface \emph{allows} feedback from any user, it appears to more successfully \emph{encourage} feedback from those whose technical knowledge or experience aligns with existing institutional norms in AI development.

\begin{table}[!t]%%%7
\centering
\tbl{Significant Associations Between User Characteristics and Feedback Behaviours\label{tab:statistical_relationships}}{
\tabcolsep3pt
\renewcommand{\arraystretch}{1.2}
\begin{tabular}{lrrrr}
\hline
\textbf{Relationship} & \textbf{$\chi^2$} & \textbf{$p$-value} & \textbf{Cramér's V} & \textbf{Interpretation} \\
\hline
Thumbs-Up/down and After Regenerating & 50.419 & <.001 & .319 & Moderate \\
Education/Training in AI and Confidence & 12.773 & .023 & .160 & Weak \\
Working in AI and Barriers & 9.821 & .023 & .143 & Weak \\
Region and Open-text feedback & 13.990 & .023 & .192 & Moderate \\
Prior AI use and Open-text feedback & 8.926 & .023 & .154 & Weak \\
Thumbs-Up/down and ChatGPT frequency & 13.689 & .045 & .166 & Weak \\
AI topics frequency and Confidence & 20.726 & .045 & .144 & Weak \\
Age and Confidence & 17.068 & .045 & .131 & Weak \\
\hline
\end{tabular}}
\end{table}

\begin{table}[!t]%%%8
\centering
\tbl{Summary of Affordances\label{tab:affordances}}{
\fontsize{9}{10}\selectfont
\tabcolsep3pt
\renewcommand{\arraystretch}{1.2}
\begin{tabular}{p{11em}<{\raggedright} l p{10em}<{\raggedright}p{13em}<{\raggedright}}
\hline
\textbf{Feature(s)} & \textbf{\makecell[l]{Mechanism of\\ Affordance}} & \textbf{Outcomes} & \textbf{For Whom/Condition} \\
\hline
Checkboxes and structure in feedback windows & \emph{Request} & Feedback formatted in the ``helpful, honest, and harmless'' framework & Feedback from users to developers (\emph{Dexterity}) \\
\hline 
Login and interaction requirement & \emph{Demand} & System usage as prerequisite for participation & Excludes non-users (\emph{Dexterity}) \\
\hline 
Thumbs-up/down buttons; Regeneration feedback options; Open-text box with categorical selection options & \emph{Encourage} & Simple, frequent feedback; Performance-focused feedback; Technical issue reporting & For users with AI/ML experience and knowledge (\emph{Cultural and institutional legitimacy}); For frequent users familiar with the system (\emph{Perception}, \emph{Dexterity}) \\
\hline 
Single-user login & \emph{Discourage} & Collective feedback; Discussion among users & Excludes collectives of users or stakeholders (\emph{Cultural and institutional legitimacy}) \\
\hline 
No edit functionality after submission; Stochasticity of LLMs & \emph{Refuse} & Modification of feedback over time; Comparative evaluation of responses & For all users (\emph{Dexterity}) \\
\hline 
Access to external platforms & \emph{Allow} & Alternative channels for feedback and discussion & For technically knowledgeable users (\emph{Dexterity}); For users with social media presence (\emph{Cultural and institutional legitimacy}) \\
\hline
\end{tabular}}
\end{table}

\subsection{Synthesis: How Affordances Shape Participation}
Table \ref{tab:affordances} outlines the affordances of the feedback features in the ChatGPT interface based on the analysis above, organised by the six mechanisms proposed by the framework. The affordances are linked to the conditions that shape how users interact with and perceive the feedback features. In summary, the interface \emph{requests} feedback formatted in the ``helpful, honest, and harmless'' framework, \emph{demands} interaction before feedback submission, and \emph{encourages} simple, frequent feedback focused on performance improvement. However, the interface \emph{discourages} collective feedback and discussions among users, while \emph{allowing} alternative channels for such discussions, and \emph{refuses} feedback modification over time. Additionally, our survey indicates that the feedback features may appeal more to users with frequent engagement, prior experience, and knowledge of AI, ML or data science.

\section{Discussion}
Participation is never fully open or anarchic; it is always formatted with ``specific rules and scripts'' \cite{Kelty2020-tt}. Feedback features in interfaces to LLMs format user input into RLHF workflows, which both legitimises user participation in the iteration of LLMs and constrains such participation to unidirectional channelling of atomised, de-contextualised user preferences. The interface design also channels user input toward easily processed feedback focused on performance-related concerns; 76\% of our respondents used thumbs-up/down feedback, but only 29\% provided open-text feedback, with technical or performance motivations significantly outweighing social or ethical concerns. This highlights a clear tension between technology companies' stated intentions (e.g., OpenAI initially sought diverse, substantive feedback to uncover potential risks and harms \cite{OpenAI2022-oj}) and how these features are actually used (simple, performance-focused feedback). While ostensibly democratising participation, we ask whether participation through feedback features is in name only, part of a ``proceduralized participation... [that] can often be implemented without any challenge to existing power relations?'' \cite{Kelty2020-tt}.

Our sample primarily comprises young, tech-savvy early adopters of ChatGPT. While this limits generalisability, the demographic skew itself highlights how affordances of feedback features in initial deployments of LLMs might privilege certain groups of users, and the narrow forms of participation we observed among these users demonstrate how the interface's affordances \emph{encouraged} simple, performance-focused feedback in this context.

\looseness-1 These constraints among early users raise broader questions about how such feedback processes might function as systems scale. If early adopters primarily provide simple, performance-focused feedback due to the interface design, users with less technical experience and knowledge, or facing other barriers, might find it even more difficult to articulate complex social or ethical concerns through these channels. This indicates the potential for a self-perpetuating dynamic where users internalise limited roles as participants in LLM iteration. This phenomenon is not unique to LLMs; similar patterns have been observed in other technological domains, such as social media\break platforms \cite{Gillespie2020-ib}.

Affordance theory as a design tool requires an orientation that ``technologies, and the worlds they build, need not be as they are and can, instead, be otherwise'' \cite{Davis2023-xk}. Drawing on affordance theory, we discuss two potential directions for the (re)design of feedback processes to move beyond the constraints revealed by our analysis, aiming at shifting participation from procedural to more substantive forms.

\subsection{Deliberative Processes}
As discussed in Section~\ref{participatoryAI}, resolving social and ethical issues in AI deployment is increasingly framed as a political task rather than a technical task. One potential direction for redesigning feedback processes to address such issues politically is to incorporate elements of deliberation---structured dialogue among users and (potentially) with developers.

Alongside aggregating citizen preferences through voting, deliberation among citizens is a key element of democratic legitimacy \cite{Fishkin1991-nv,Knight1994-lc,Curato2017-iw}. Deliberative democracy involves an association of members who govern their affairs through deliberation among the members \cite{Elster1998-vl}---making collective decisions based not only on their preferences but also on their reasoning \cite{Cohen2005-af}. Such associations typically involve broad groups of people and encourage participants to consider others' perspectives and potentially revise their own views over time

Deliberative processes could change the existing affordances of user participation in iterating LLMs. For example, where current interfaces \emph{discourage} reasoned discussion among users, deliberative platforms could explicitly \emph{request} reasoning from participants and \emph{allow} for bidirectional communication between users and developers, along with evolving perspectives, moving toward genuine feedback loops rather than the unidirectional channels currently prevalent. Interestingly, some respondents identified several other platforms on which they discuss their experiences and preferences regarding ChatGPT, engaging in the kind of collective deliberation that the existing feedback features \emph{discourage.} For example, OpenAI discussion forums \emph{allow} communication between users and developers and subreddits such as r/ChatGPT \emph{allow} discussion among users (without monitoring or oversight by OpenAI). However, it is unclear whether or how these discussions feed back into development.

Some technology companies have been experimenting with deliberative modes for feedback about AI systems. Recent efforts by OpenAI to fund research into technical solutions for ``democratic inputs into AI'' \cite{Eloundou2024-xa} and experiments by Anthropic to draft a constitution for Claude through deliberative processes \cite{Huang2024-zs} have attracted significant attention. These efforts go beyond eliciting and aggregating user preferences. They encourage deliberation among users and (potentially) other stakeholders, while demonstrating how multiple participants' preferences and reasoning feed into development workflows. In addition, Meta has experimented with Community Forums to deliberate on the rulesets for Generative AI \cite{Broxmeyer2024-zw}, and Google has developed the STELA process for deliberative discussions with historically underrepresented groups in the U.S.A. to understand their priorities and concerns with respect to model outputs \cite{Bergman2024-yo}.

While these processes address some of the limitations of existing feedback features, they remain largely experimental and peripheral to the dominant method of feedback submission through browser-based interfaces. Most of these approaches (with the STELA process as an exception) take as a given that consensus on the output of the process is desirable and achievable. However, refusal, agonism and dissensus are all key elements of democratic processes \cite{DiSalvo2015-dd, DiSalvo2010-uq, Zong2023-fe}. When consulting multiple and diverse communities, values and preferences for AI design may be in conflict \cite{Sambasivan2021-il,Qadri2023-fv} and difficult to reconcile \cite{Santurkar2023-tc,Durmus2023-ot,Miller2007-mm}. Such difficulty is particularly acute where the processes are managed by those with commercial interests in consensus outcomes, reflecting long-standing concerns in participatory design \cite{Sanders2014-kh}.

Moreover, such democratic inputs to the alignment problem, much like the existing feedback features they are intended to augment or replace, narrowly scope participation around the output of specific families of models, \emph{refusing} participants meaningful engagement with earlier development stages \cite{Delgado2023-tm}. Although potentially addressing some limitations of current feedback features, users and stakeholders remain as consumers rather than co-creators.

\subsection{Infrastructuring Participation}
While our insights are based on early adopters, the constraints revealed by our analysis highlight the limitations of current processes for feedback on harmful model outputs or risks from LLM deployment. Existing processes appear to be efficient channels for atomised preferences from specific user groups, but insufficient conduits for dialogue on social and ethical issues regarding LLMs. Echoing Arnstein's \cite{Arnstein1969-ku} critique of participation in the context of urban planning, identifying problems in the proceduralisation of participation through existing processes is not simply a call for more involvement, but instead requires a fundamental rethinking of the process.

\looseness-1 We advocate for moves beyond aligning model outputs to specific user preferences toward feedback processes that embrace productive disagreement and plurality. As DiSalvo \cite{DiSalvo2010-uq} argues, designers can enhance democratic engagement by ``creating and enabling ... spaces of contest'' where ``difference and dissensus are brought forward and the assumptions and actions that shape power relations and influence are revealed and challenged'' \cite{DiSalvo2010-uq}. To create feedback channels capable of supporting these more contested forms of engagement requires actively ``infrastructuring'' participation---creating and sustaining social, technical, and institutional structures with different affordances to address matters of concern to participants \cite{Le_Dantec2013-zu}. Such infrastructure must aim at changing the conditions of participation, to actively \emph{encourage} engagement from stakeholders affected by LLM deployments, including non-users and those without technical backgrounds, and \emph{allow} forms of input that address participants' concerns beyond immediate system\break performance.

Recent research offers promising practical directions. The STELA process \cite{Bergman2024-yo} demonstrates one method for conducting community-centred deliberative discussions about LLMs with marginalised groups who may or may not be users of LLMs, albeit focused on aligning models with the preferences of those groups of people. \citet{Suresh2024-vt} expand on this concept by proposing a comprehensive framework for domain-specific community infrastructure, including shared technical resources, norms, and governance structures. Their approach aims at enabling participation by stakeholder groups and institutions to shape the development and application of foundation models in specific domains. Building on these ideas, we encourage participatory design workshops with stakeholders, in fields like education, healthcare, and journalism, that use techniques encouraging agonism (see, e.g., \cite{Bjorgvinsson2012Agonistic}) to surface conflicts and refine such domain-specific infrastructure.

Building and sustaining this infrastructure is not a one-off task; it requires ongoing effort to secure resources, build trust with participants, navigate conflict, ensure local relevance while extending across domains and regions, and translate outputs into actionable development inputs \cite{Le_Dantec2013-zu}. However, these are not reasons to default to simpler, constrained modes of participation. If the goal of feedback processes is to afford meaningful public influence over technologies with profound societal impact, then the focus of research and effort must be on addressing such challenges to infrastructuring sustained, pluralistic, and sometimes contentious dialogue with affected stakeholders.

\section{Conclusion}
Our analysis of ChatGPT's feedback features reveals significant constraints on user participation, based on a survey of early adopters' experiences. The interface appears to \emph{encourage} simple, frequent, and performance-focused feedback while \emph{discouraging} collective input and discussions among users. We argue that this constrained approach to participation suggests limitations for meaningful dialogue about social and ethical issues regarding LLM deployment as system usage becomes more widespread.

Three research directions could substantiate and extend our analysis. Firstly, empirical studies could test whether deliberative processes better capture social and ethical concerns than current processes. Additionally, longitudinal research could track both how feedback patterns evolve as broader populations adopt LLMs, and how organisations frame their feedback processes over time. Such work might reveal growing divergences between companies' stated commitments and actual practices. Most importantly, we encourage ethnographic and action research examining the practical work of infrastructuring. Ethnographic studies could document ongoing attempts to create participatory structures in specific domains, while action researchers could collaborate with affected communities to prototype and sustain new forms of engagement (see, e.g., \cite{Tseng2025Ownership}).

Without fundamental changes, the scope of user (and non-user) input will remain constrained. To move beyond perfunctory engagement requires shifting focus from user preference elicitation to building participatory infrastructures with affordances that \emph{allow} and \emph{encourage} sustained dialogue---including disagreement---with affected stakeholders about the purpose and applications of LLMs. As these technologies become increasingly embedded in information, workplace, and entertainment services, the urgency of this task cannot be overstated.

\begin{acks}
This research is supported by an Australian Government \textbf{Research Training Program (RTP)} Scholarship and a Florence Violet McKenzie Scholarship. We thank the organisers and participants of the CHI 2023 Workshop on Designing Technology and Policy Simultaneously and the 2023 AusSTS Conference for feedback on earlier drafts. We also thank Professor Alice Richardson of the Statistical Support Network, ANU, for helpful advice on the statistical analysis in this article. Finally, we thank our reviewers for their feedback and suggestions.
\end{acks}

\newpage

\appendix
\section*{Appendices}
\section{Survey Questions}

Appendix A lists the substantive questions from the online survey instrument used for this study. 

\subsection*{Screening}
\begin{enumerate}
  \item Have you used ChatGPT in any capacity? e.g., to generate text, answer a question, or simply experiment with the system.
  \begin{itemize}
    \item[\mdash] Yes
    \item[\mdash] No
  \end{itemize}
\end{enumerate}

\subsection*{ChatGPT Usage}
\begin{enumerate}
  \setcounter{enumi}{1} % Start numbering from 2
  \item How often have you used ChatGPT since its release in November 2022?
  \begin{itemize}
    \item[\mdash] Daily
    \item[\mdash] A few times a week
    \item[\mdash] Once a week
    \item[\mdash] Monthly
    \item[\mdash] Rarely
  \end{itemize}

  \item In which of the following ways have you used ChatGPT? Please select all that apply.
  \begin{itemize}
    \item[\mdash] Academic research assistance (e.g., generating citations, summarising articles)
    \item[\mdash] Generating text (e.g., drafting e-mails, writing essays or reports)
    \item[\mdash] Answering specific questions (e.g., factual, technical, or theoretical questions)
    \item[\mdash] Creative writing (e.g., stories, poetry, or brainstorming ideas)
    \item[\mdash] Language learning or translation
    \item[\mdash] Personal advice or emotional support (e.g., relationship or career advice)
    \item[\mdash] Professional or technical problem-solving (e.g., programming, troubleshooting)
    \item[\mdash] Entertainment or casual conversation (e.g., jokes, trivia, or random facts)
    \item[\mdash] Other (Specify) % Added Specify note
  \end{itemize}

  \item In which ways have you used ChatGPT most frequently? Please order from 1 = most frequent, 2 = next most frequent, and so on.
  \begin{itemize}
    \item[\mdash] Academic research assistance (e.g., generating citations, summarising articles)
    \item[\mdash] Generating text (e.g., drafting e-mails, writing essays or reports)
    \item[\mdash] Answering specific questions (e.g., factual, technical, or theoretical questions)
    \item[\mdash] Creative writing (e.g., stories, poetry, or brainstorming ideas)
    \item[\mdash] Language learning or translation
    \item[\mdash] Personal advice or emotional support (e.g., relationship or career advice)
    \item[\mdash] Professional or technical problem-solving (e.g., programming, troubleshooting)
    \item[\mdash] Entertainment or casual conversation (e.g., jokes, trivia, or random facts)
    \item[\mdash] Other (Specify) % Added Specify note
  \end{itemize}
\end{enumerate}

\subsection*{Feedback Submission: Thumbs}
\begin{enumerate}
  \setcounter{enumi}{4} % Start numbering from 5
  \item Have you submitted feedback on ChatGPT's responses using the up and down thumbs?
  \begin{itemize}
    \item[\mdash] Yes
    \item[\mdash] No
  \end{itemize}

  \item (If Yes to previous) How often have you submitted feedback using the up and down thumbs?
  \begin{itemize}
    \item[\mdash] Every time I have used ChatGPT
    \item[\mdash] Most times I have used ChatGPT
    \item[\mdash] About half the time I have used ChatGPT
    \item[\mdash] Some times I have used ChatGPT
    \item[\mdash] Rarely
  \end{itemize}
\end{enumerate}

\subsection*{Feedback Submission: Open-Text}
\begin{enumerate}
  \setcounter{enumi}{6} % Start numbering from 7
  \item Have you submitted open-text feedback on ChatGPT's responses?
  \begin{itemize}
    \item[\mdash] Yes
    \item[\mdash] No
  \end{itemize}

  \item (If Yes to previous) How often have you submitted open-text feedback?
  \begin{itemize}
    \item[\mdash] Every time I have used ChatGPT
    \item[\mdash] Most times I have used ChatGPT
    \item[\mdash] About half the time I have used ChatGPT
    \item[\mdash] Some times I have used ChatGPT
    \item[\mdash] Rarely
  \end{itemize}
\end{enumerate}

\subsection*{Feedback Submission: Regenerate Response}
\begin{enumerate}
  \setcounter{enumi}{8} % Start numbering from 9
  \item Have you submitted feedback after regenerating one of ChatGPT's responses?
  \begin{itemize}
    \item[\mdash] Yes
    \item[\mdash] No
  \end{itemize}

  \item (If Yes to previous) How often have you submitted feedback after regenerating a response?
  \begin{itemize}
    \item[\mdash] Every time I have regenerated a response
    \item[\mdash] Most times I have regenerated a response
    \item[\mdash] About half the time I have regenerated a response
    \item[\mdash] Some times I have regenerated a response
    \item[\mdash] Rarely
  \end{itemize}
\end{enumerate}

\subsection*{Reasons for Not Submitting Feedback}
\begin{enumerate}
  \setcounter{enumi}{10} % Start numbering from 11
  \item (If never submitted any feedback) What is the main reason you have not submitted feedback through the ChatGPT interface? Please select the main reason only. NB: `Performance' refers to the helpfulness, truthfulness, or harmlessness of responses from ChatGPT to a prompt.
  \begin{itemize}
    \item[\mdash] Unaware of feedback option
    \item[\mdash] Satisfied with the performance and have not felt the need to submit feedback
    \item[\mdash] Unsatisfied with the performance, but do not believe feedback will lead to improvements
    \item[\mdash] Unsure how to provide feedback
    \item[\mdash] Not enough time or interest in submitting feedback
    \item[\mdash] Other (Specify) % Added Specify note
  \end{itemize}
\end{enumerate}

\subsection*{Motivations}
\begin{enumerate}
  \setcounter{enumi}{11} % Start numbering from 12
  \item (If submitted open-text feedback) Think back to the most recent time you submitted feedback through the interface. What motivated you to submit feedback about ChatGPT? Please select all that apply.
  \begin{itemize}
    \item[\mdash] To report a technical issue or bug
    \item[\mdash] To suggest new features or improvements
    \item[\mdash] To express social or ethical concerns about ChatGPT
    \item[\mdash] To comment on use-case specific feedback
    \item[\mdash] To contribute to AI development
    \item[\mdash] Other (Specify) % Added Specify note
  \end{itemize}

  \item (If selected multiple motivations previously) What was your main motivation for submitting feedback? Please place a 1 in the box next to your main motivation.
   \begin{itemize}
    \item[\mdash] To report a technical issue or bug
    \item[\mdash] To suggest new features or improvements
    \item[\mdash] To express social or ethical concerns about ChatGPT
    \item[\mdash] To comment on use-case specific feedback
    \item[\mdash] To contribute to AI development
    \item[\mdash] Other (Specify) % Added Specify note
  \end{itemize}

  \item (If submitted open-text feedback) What kind of feedback do you think is most useful for improving ChatGPT's performance or impact? e.g., reporting technical issues, suggesting new features, expressing social or ethical concerns, and so on. ``Performance'' refers to the helpfulness, truthfulness, or harmlessness of responses from ChatGPT to a prompt. ``Impact'' refers to changes in society and culture from ChatGPT. [Open Text] % Added marker

  \item (If submitted open-text feedback) Have you received any follow-up communication or updates regarding your feedback?\enlargethispage{-8pt}
  \begin{itemize}
    \item[\mdash] Yes
    \item[\mdash] No
  \end{itemize}

  \item (If Yes to previous) How satisfied are you with the response or resolution?
  \begin{itemize}
    \item[\mdash] Extremely satisfied
    \item[\mdash] Somewhat satisfied
    \item[\mdash] Neither satisfied nor dissatisfied
    \item[\mdash] Somewhat dissatisfied
    \item[\mdash] Extremely dissatisfied
  \end{itemize}

  \item (If submitted open-text feedback) How much do you agree with this statement: ``My reasons for submitting feedback about ChatGPT through the interface have been addressed by OpenAI.''
  \begin{itemize}
    \item[\mdash] Strongly agree
    \item[\mdash] Somewhat agree
    \item[\mdash] Neither agree nor disagree
    \item[\mdash] Somewhat disagree
    \item[\mdash] Strongly disagree
  \end{itemize}
\end{enumerate}

\subsection*{Reflections}
\begin{enumerate}
  \setcounter{enumi}{17} % Start numbering from 18
  \item (If submitted any feedback) Was submitting feedback through the ChatGPT interface easy or difficult? % Added condition
  \begin{itemize}
    \item[\mdash] Extremely easy
    \item[\mdash] Somewhat easy
    \item[\mdash] Neither easy nor difficult
    \item[\mdash] Somewhat difficult
    \item[\mdash] Extremely difficult
  \end{itemize}

  \item How likely are you to submit feedback through the ChatGPT interface in the future?
  \begin{itemize}
    \item[\mdash] Extremely likely
    \item[\mdash] Somewhat likely
    \item[\mdash] Neither likely nor unlikely
    \item[\mdash] Somewhat unlikely
    \item[\mdash] Extremely unlikely
  \end{itemize}
  
  \item (Conditional logic based on Q19) Why are you likely to submit feedback through the ChatGPT interface in the future? [Open Text] % Added marker

  \item (Conditional logic based on Q19) Why are you not likely to submit feedback through the ChatGPT interface in the future? [Open Text] % Added marker
\end{enumerate}

\subsection*{Feedback via Other Channels}
\begin{enumerate}
  \setcounter{enumi}{21} % Start numbering from 22
  \item Have you provided feedback on ChatGPT's performance or impact through other channels or platforms? e.g., social media platforms, online discussion forums, contacting OpenAI, and so on. ``Performance'' refers to the helpfulness, truthfulness, or harmlessness of responses from ChatGPT to a prompt. ``Impact'' refers to changes in society and culture from ChatGPT.
  \begin{itemize}
    \item[\mdash] Yes
    \item[\mdash] No
  \end{itemize}

  \item (If Yes to previous) Which other channels or platforms have you used to provide feedback about ChatGPT's performance or impact? Please select all that apply.
  \begin{itemize}
    \item[\mdash] X (Please list any specific \# or communities) % Allows text input
    \item[\mdash] Mastodon (Please list any specific \# or servers) % Allows text input
    \item[\mdash] Linkedin (Please list any groups) % Allows text input
    \item[\mdash] Facebook (Please list any groups) % Allows text input
    \item[\mdash] Reddit forums (Please specify which subreddits e.g., r/ChatGPT) % Allows text input
    \item[\mdash] Discord servers (Please specify which servers) % Allows text input
    \item[\mdash] Other online discussion forums (Please specify e.g., Github Community, OpenAI Community) % Allows text input
    \item[\mdash] Contacting OpenAI (e.g., via e-mail)
    \item[\mdash] Other (Specify) % Added Specify note
  \end{itemize}

  \item (If Yes to using other channels) How likely are you to use those channels or platforms to provide feedback on ChatGPT in the future?
  \begin{itemize}
    \item[\mdash] Extremely likely
    \item[\mdash] Somewhat likely
    \item[\mdash] Neither likely nor unlikely
    \item[\mdash] Somewhat unlikely
    \item[\mdash] Extremely unlikely
  \end{itemize}

  \item (Conditional logic based on Q24) Why are you likely to provide feedback on ChatGPT through those channels or platforms in the future? [Open Text] % Added marker

  \item (Conditional logic based on Q24) Why are you not likely to provide feedback on ChatGPT through those channels or platforms in the future? [Open Text] % Added marker
\end{enumerate}

\subsection*{Demographics}
\begin{enumerate}
  \setcounter{enumi}{26} % Start numbering from 27
  \item What is your age?
  \begin{itemize}
    \item[\mdash] Under 18
    \item[\mdash] 18-24
    \item[\mdash] 25-34
    \item[\mdash] 35-44
    \item[\mdash] 45-54
    \item[\mdash] 55-64
    \item[\mdash] 65+
    \item[\mdash] Prefer not to say
  \end{itemize}

  \item How would you describe your gender?
  \begin{itemize}
    \item[\mdash] Man
    \item[\mdash] Woman
    \item[\mdash] Non-binary/third gender
    \item[\mdash] Prefer to self-describe (Specify) % Added Specify note
    \item[\mdash] Prefer not to say
  \end{itemize}

  \item In which country do you currently reside? [Open Text - Likely] % Added marker

  \item What are the main languages you speak daily? Select up to three languages. [Open Text] % Added marker

  \item Did you experience any barriers accessing ChatGPT or submitting feedback about ChatGPT?
  \begin{itemize}
    \item[\mdash] Computer hardware barriers
    \item[\mdash] Internet connectivity barriers
    \item[\mdash] Health and medical barriers
    \item[\mdash] Language barriers
    \item[\mdash] Other (Specify) % Added Specify note
    \item[\mdash] None of the above
  \end{itemize}

  \item How often do you engage with or read about AI-related topics? e.g., news articles, research papers, podcasts, and so on.\enlargethispage{9pt}
  \begin{itemize}
    \item[\mdash] Daily
    \item[\mdash] Weekly
    \item[\mdash] Monthly
    \item[\mdash] Rarely
    \item[\mdash] Never
  \end{itemize}

  \item Before using ChatGPT, have you used any AI, machine learning, or data science tools, beyond basic interaction or exploration? Please select ``Yes'' if you have used AI, machine learning, or data science tools for a project, work, or in-depth exploration.
  \begin{itemize}
    \item[\mdash] Yes
    \item[\mdash] No
  \end{itemize}

  \item Have you completed any formal education, training, or coursework related to AI, machine learning, or data science?
  \begin{itemize}
    \item[\mdash] Yes
    \item[\mdash] No
  \end{itemize}

  \item Are you currently working in, or have you ever worked in, a field related to AI, machine learning, or data science?
  \begin{itemize}
    \item[\mdash] Yes
    \item[\mdash] No
  \end{itemize}

  \item How much do you agree with the following statement: ``I feel confident in my ability to evaluate and provide feedback on the performance and/or impact of generative A systems, such as ChatGPT''?
  \begin{itemize}
    \item[\mdash] Strongly agree
    \item[\mdash] Somewhat agree
    \item[\mdash] Neither agree nor disagree
    \item[\mdash] Somewhat disagree
    \item[\mdash] Strongly disagree
  \end{itemize}
\end{enumerate}

\clearpage

\section{Survey Results}
Appendix B compiles the survey results from this study.

\begin{table}[!h]%%%9
\centering
\tbl{Frequency of ChatGPT Usage\label{tab:chatgpt-usage-frequency}}{
\renewcommand{\arraystretch}{1.05}
\begin{tabular}{lrr}
\hline
Frequency & Count & Percentage \\
\hline
Daily & 64 & 12.2\% \\
A few times a week & 216 & 41.1\% \\
Once a week & 96 & 18.3\% \\
Monthly & 84 & 16.0\% \\
Rarely & 65 & 12.4\% \\
\hline
Total & 525 & 100.0\% \\
\hline
\end{tabular}}
\vspace*{2pt}
\tabnote{\hangindent.5em\ \ (Response rate: 525/526).}
\end{table}

\begin{table}[!h]%%%10
\centering
\tbl{Ways of Using ChatGPT\label{tab:chatgpt-usage-ways}}{
\renewcommand{\arraystretch}{1.05}
\begin{tabular}{lrr}
\hline
Usage & Count & Percentage \\
\hline
Answering specific questions & 375 & 71.3\% \\
Generating text & 332 & 62.9\% \\
Academic research assistance & 268 & 51.0\% \\
Entertainment or casual conversation & 226 & 43.0\% \\
Professional or technical problem-solving & 222 & 42.2\% \\
Creative writing & 214 & 40.7\% \\
Language learning or translation & 122 & 23.3\% \\
Personal advice or emotional support & 109 & 20.7\% \\
Other & 22 & 4.2\% \\
\hline
\end{tabular}}
\vspace*{2pt}
\tabnote{\hangindent.5em\ \ (Response rate: 526/526, Multiple selections allowed).}
\end{table}

\begin{table}[!h]%%%11
\centering
\tbl{Submitted Thumbs-Up/Down Feedback\label{tab:thumbs-feedback}}{
\renewcommand{\arraystretch}{1.05}
\begin{tabular}{lrr}
\hline
Submitted feedback & Count & Percentage \\
\hline
Yes & 400 & 76.0\% \\
No & 126 & 24.0\% \\
\hline
Total & 526 & 100.0\% \\
\hline
\end{tabular}}
\vspace*{2pt}
\tabnote{\hangindent.5em\ \ (Response rate: 526/526).}
\end{table}

\begin{table}[!t]%%%12
\centering
\tbl{Frequency of Thumbs-Up/Down Feedback\label{tab:thumbs-feedback-frequency}}{
\renewcommand{\arraystretch}{1.05}
\begin{tabular}{lrr}
\hline
Frequency & Count & Percentage \\
\hline
Every time I have used ChatGPT & 23 & 5.8\% \\
Most times I have used ChatGPT & 92 & 23.0\% \\
About half the time I have used ChatGPT & 98 & 24.5\% \\
Sometimes I have used ChatGPT & 139 & 34.8\% \\
Rarely & 48 & 12.0\% \\
\hline
Total & 400 & 100.0\% \\
\hline
\end{tabular}}
\vspace*{2pt}
\tabnote{\hangindent.5em\ \ (Response rate: 400/400).}
\end{table}

\begin{table}[!t]%%%13
\centering
\tbl{Submitted Open-Text Feedback\label{tab:open-text-feedback}}{
\renewcommand{\arraystretch}{1.1}
\begin{tabular}{lrr}
\hline
Submitted feedback & Count & Percentage \\
\hline
Yes & 152 & 29.0\% \\
No & 248 & 47.1\% \\
N/A & 126 & 23.9\% \\
\hline
Total & 526 & 100.0\% \\
\hline
\end{tabular}}
\vspace*{2pt}
\tabnote{\hangindent.5em\ \ (Response rate: 526/526, N/A = ``No'' to submitting thumbs-up/down feedback.)}
\end{table}

\begin{table}[!t]%%%14
\centering
\tbl{Submitted Open-Text Feedback (``Yes'' to thumbs-up/down feedback)\label{tab:open-text-feedback-yes}}{
\renewcommand{\arraystretch}{1.1}
\begin{tabular}{lrr}
\hline
Submitted feedback & Count & Percentage \\
\hline
Yes & 152 & 38.0\% \\
No & 248 & 62.0\% \\
\hline
Total & 400 & 100.0\% \\
\hline
\end{tabular}}
\vspace*{2pt}
\tabnote{\hangindent.5em\ \ (Response rate: 400/400).}
\end{table}

\begin{table}[!t]%%%15
\centering
\tbl{Frequency of Open-Text Feedback\label{tab:open-text-feedback-frequency}}{
\renewcommand{\arraystretch}{1.1}
\begin{tabular}{lrr}
\hline
Frequency & Count & Percentage \\
\hline
Every time I have used ChatGPT & 3 & 2.0\% \\
Most times I have used ChatGPT & 20 & 13.2\% \\
About half the time I have used ChatGPT & 17 & 11.2\% \\
Sometimes I have used ChatGPT & 77 & 50.7\% \\
Rarely & 35 & 23.0\% \\
\hline
Total & 152 & 100.0\% \\
\hline
\end{tabular}}
\vspace*{2pt}
\tabnote{\hangindent.5em\ \ (Response rate: 152/152).}
\end{table}

\begin{table}[!t]%%%16
\centering
\tbl{Submitted Feedback After Regenerating a Response\label{tab:regenerate-feedback}}{
\renewcommand{\arraystretch}{1.1}
\begin{tabular}{lrr}
\hline
Submitted feedback & Count & Percentage \\
\hline
Yes & 331 & 62.9\% \\
No & 195 & 37.1\% \\
\hline
Total & 526 & 100.0\% \\
\hline
\end{tabular}}
\vspace*{2pt}
\tabnote{\hangindent.5em\ \ (Response rate: 526/526).}
\end{table}

\begin{table}[!t]%%%17
\centering
\tbl{Frequency of Feedback After Regenerating a Response\label{tab:regenerate-feedback-frequency}}{
\renewcommand{\arraystretch}{1.1}
\begin{tabular}{lrr}
\hline
Frequency & Count & Percentage \\
\hline
Every time I have regenerated a response & 25 & 7.6\% \\
Most times I have regenerated a response & 67 & 20.2\% \\
About half the time I have regenerated a response & 65 & 19.6\% \\
Sometimes I have regenerated a response & 120 & 36.3\% \\
Rarely & 54 & 16.3\% \\
\hline
Total & 331 & 100.0\% \\
\hline
\end{tabular}}
\vspace*{2pt}
\tabnote{\hangindent.5em\ \ (Response rate: 331/331).}
\end{table}

\begin{table}[!t]%%%18
\centering
\tbl{Reasons for Not Submitting Feedback\label{tab:no-feedback-reasons}}{
\fontsize{8.5}{9.5}\selectfont
\tabcolsep3pt
\renewcommand{\arraystretch}{1.2}
\begin{tabular}{lrr}
\hline
Reasons & Count & Percentage \\
\hline
Unaware of feedback option & 27 & 34.6\% \\
Not enough time or interest in submitting feedback & 21 & 26.9\% \\
Satisfied with the performance and have not felt need to submit feedback & 17 & 21.8\% \\
Unsure how to provide feedback & 7 & 9.0\% \\
Other & 5 & 6.4\% \\
Unsatisfied with the performance, but do not believe feedback will lead to improvements & 1 & 1.3\% \\
\hline
Total & 78 & 100.0\% \\
\hline
\end{tabular}}
\vspace*{2pt}
\tabnote{\hangindent.5em\ \ (Response rate: 78/78).}
\end{table}

\begin{table}[!t]%%%19
\centering
\tbl{Motivations for Providing Feedback\label{tab:feedback-motivations}}{
\renewcommand{\arraystretch}{1.1}
\begin{tabular}{lr}
\hline
Motivation & Weighted frequency \\
\hline
To contribute to AI development & 159 \\
To comment on use-case specific feedback & 94 \\
To report a technical issue or bug & 74 \\
To suggest new features or improvements & 73 \\
To express social or ethical concerns about ChatGPT & 32 \\
Other & 15 \\
\hline
\end{tabular}}
\vspace*{2pt}
\tabnote{\hangindent.5em\ \ (Response rate: 152/152).}
\vspace*{2pt}
\tabnote{\hangindent.5em\ \ \textbf{Note:} Weighted frequencies based on a two-part question. Part A (multiple selection) responses weighted as 1; Part B (main motivation) responses weighted as 2.}
\end{table}

\begin{table}[!t]%%%20
\centering
\tbl{Received Follow-up Regarding Feedback\label{tab:feedback-followup}}{
\renewcommand{\arraystretch}{1.1}
\begin{tabular}{lrr}
\hline
Response & Count & Percentage \\
\hline
Yes & 17 & 11.2\% \\
No & 135 & 88.8\% \\
\hline
Total & 152 & 100.0\% \\
\hline
\end{tabular}}
\vspace*{2pt}
\tabnote{\hangindent.5em\ \ (Response rate: 152/152).}
\end{table}

\begin{table}[!t]%%%21
\centering
\tbl{Satisfaction with Follow-up\label{tab:followup-satisfaction}}{
\renewcommand{\arraystretch}{1.1}
\begin{tabular}{lrr}
\hline
Satisfaction level & Count & Percentage \\
\hline
Extremely satisfied & 5 & 29.4\% \\
Somewhat satisfied & 10 & 58.8\% \\
Neither satisfied nor dissatisfied & 2 & 11.8\% \\
Somewhat dissatisfied & 0 & 0\% \\
Extremely dissatisfied & 0 & 0\% \\
\hline
Total & 17 & 100.0\% \\
\hline
\end{tabular}}
\vspace*{2pt}
\tabnote{\hangindent.5em\ \ (Response rate: 17/17).}
\end{table}

\begin{table}[!t]%%%22
\centering
\tbl{Reasons for Submitting Feedback Addressed\label{tab:feedback-reasons-addressed}}{
\renewcommand{\arraystretch}{1.1}
\begin{tabular}{lrr}
\hline
Agreement level & Count & Percentage \\
\hline
Strongly agree & 8 & 5.3\% \\
Somewhat agree & 36 & 23.7\% \\
Neither agree nor disagree & 72 & 47.4\% \\
Somewhat disagree & 25 & 16.4\% \\
Strongly disagree & 11 & 7.2\% \\
\hline
Total & 152 & 100.0\% \\
\hline
\end{tabular}}
\vspace*{2pt}
\tabnote{\hangindent.5em\ \ (Response rate: 152/152).}
\end{table}

\begin{table}[!t]%%%23
\centering
\tbl{Ease of Submitting Feedback\label{tab:feedback-ease}}{
\renewcommand{\arraystretch}{1.1}
\begin{tabular}{lrr}
\hline
Level of ease & Count & Percentage \\
\hline
Extremely easy & 292 & 66.1\% \\
Somewhat easy & 128 & 29.0\% \\
Neither easy nor difficult & 18 & 4.1\% \\
Somewhat difficult & 4 & 0.9\% \\
Extremely difficult & 0 & 0\% \\
\hline
Total & 442 & 100.0\% \\
\hline
\end{tabular}}
\vspace*{2pt}
\tabnote{\hangindent.5em\ \ (Response rate: 442/443).}
\end{table}

\begin{table}[!t]%%%24
\centering
\tbl{Likelihood of Submitting Feedback in Future\label{tab:future-feedback-likelihood}}{
\renewcommand{\arraystretch}{1.1}
\begin{tabular}{lrr}
\hline
Likelihood & Count & Percentage \\
\hline
Extremely likely & 161 & 30.7\% \\
Somewhat likely & 255 & 48.7\% \\
Neither likely nor unlikely & 67 & 12.8\% \\
Somewhat unlikely & 34 & 6.5\% \\
Extremely unlikely & 7 & 1.3\% \\
\hline
Total & 524 & 100.0\% \\
\hline
\end{tabular}}
\vspace*{2pt}
\tabnote{\hangindent.5em\ \ (Response rate: 524/526).}
\end{table}

\begin{table}[!t]%%%25
\centering
\tbl{Provided Feedback via Other Platforms or Channels\label{tab:other-feedback-channels}}{
\renewcommand{\arraystretch}{1.1}
\begin{tabular}{lrr}
\hline
Provided feedback & Count & Percentage \\
\hline
Yes & 103 & 19.6\% \\
No & 423 & 80.4\% \\
\hline
Total & 526 & 100.0\% \\
\hline
\end{tabular}}
\vspace*{2pt}
\tabnote{\hangindent.5em\ \ (Response rate: 526/526).}
\end{table}

\begin{table}[!t]%%%26
\centering
\tbl{Other Channels or Platforms\label{tab:other-feedback-platforms}}{
\renewcommand{\arraystretch}{1.07}
\begin{tabular}{lrr}
\hline
Channels & Count & Percentage \\
\hline
X/Twitter & 42 & 40.8\% \\
Facebook & 18 & 17.5\% \\
Other & 18 & 17.5\% \\
Reddit forums & 17 & 16.5\% \\
Discord servers & 17 & 16.5\% \\
LinkedIn & 10 & 9.7\% \\
Contacting OpenAI (e.g., via e-mail) & 10 & 9.7\% \\
Other online discussion forums & 9 & 8.7\% \\
Mastodon & 3 & 2.9\% \\
\hline
\end{tabular}}
\vspace*{2pt}
\tabnote{\hangindent.5em\ \ (Response rate 103/103, Multiple selections allowed).}
\end{table}

\begin{table}[!t]%%%27
\centering
\tbl{Likelihood of Submitting Feedback via Other Channels or Platforms in Future\label{tab:other-channels-future-likelihood}}{
\renewcommand{\arraystretch}{1.07}
\begin{tabular}{lrr}
\hline
Likelihood & Count & Percentage \\
\hline
Extremely likely & 28 & 27.2\% \\
Somewhat likely & 46 & 44.7\% \\
Neither likely nor unlikely & 15 & 14.6\% \\
Somewhat unlikely & 10 & 9.7\% \\
Extremely unlikely & 4 & 3.9\% \\
\hline
Total & 103 & 100.0\% \\
\hline
\end{tabular}}
\vspace*{2pt}
\tabnote{\hangindent.5em\ \ (Response rate: 103/103).}
\end{table}

\begin{table}[!t]%%%28
\centering
\tbl{Age\label{tab:age}}{
\renewcommand{\arraystretch}{1.07}
\begin{tabular}{lrr}
\hline
Age & Count & Percentage \\
\hline
18-24 & 222 & 42.4\% \\
25-34 & 184 & 35.1\% \\
35-44 & 72 & 13.7\% \\
45-54 & 31 & 5.9\% \\
55-64 & 8 & 1.5\% \\
65+ & 4 & 0.8\% \\
Prefer not to say & 3 & 0.6\% \\
\hline
Total & 524 & 100.0\% \\
\hline
\end{tabular}}
\vspace*{2pt}
\tabnote{\hangindent.5em\ \ (Response rate: 524/526).}
\end{table}

\begin{table}[!t]%%%29
\centering
\tbl{Gender\label{tab:gender}}{
\renewcommand{\arraystretch}{1.07}
\begin{tabular}{lrr}
\hline
Gender & Count & Percentage \\
\hline
Man & 329 & 63.1\% \\
Woman & 182 & 34.9\% \\
Non-binary / third gender & 6 & 1.2\% \\
Prefer not to say & 3 & 0.6\% \\
Prefer to self-describe & 1 & 0.2\% \\
\hline
Total & 521 & 100.0\% \\
\hline
\end{tabular}}
\vspace*{2pt}
\tabnote{\hangindent.5em\ \ (Response rate: 521/526).}
%%\vspace*{10pt}
\end{table}

\begin{table}[!t]%%%30
\centering
\tbl{Region\label{tab:region}}{
\renewcommand{\arraystretch}{1.1}
\begin{tabular}{lrr}
\hline
Region & Count & Percentage \\
\hline
Europe & 281 & 55.4\% \\
Africa & 79 & 15.6\% \\
United Kingdom & 52 & 10.3\% \\
Oceania & 33 & 6.5\% \\
North America & 23 & 4.5\% \\
Central America & 21 & 4.1\% \\
South America & 9 & 1.8\% \\
Asia & 9 & 1.8\% \\
\hline
Total & 507 & 100.0\% \\
\hline
\end{tabular}}
\vspace*{2pt}
\tabnote{\hangindent.5em\ \ (Response rate: 507/526).}
\end{table}

\begin{table}[!t]%%%31
\centering
\tbl{Languages Spoken Daily (Grouped)\label{tab:languages-grouped}}{
\tabcolsep8pt
\renewcommand{\arraystretch}{1.1}
\begin{tabular}{lrr}
\hline
Languages & Count & Percentage \\
\hline
English (UK) & 163 & 32.0\% \\
English (US) & 116 & 22.8\% \\
Portuguese & 68 & 13.4\% \\
Polish & 62 & 12.2\% \\
Italian & 33 & 6.5\% \\
English (AU) & 32 & 6.3\% \\
Zulu & 29 & 5.7\% \\
Spanish (including Latin American) & 49 & 9.6\% \\
Greek & 15 & 2.9\% \\
German (Standard and Swiss) & 14 & 2.8\% \\
Sotho & 13 & 2.6\% \\
French (including Canadian) & 9 & 1.8\% \\
Other African languages & 8 & 1.6\% \\
Dutch & 6 & 1.2\% \\
Czech & 5 & 1.0\% \\
Other European languages & 14 & 2.8\% \\
Other Asian languages & 11 & 2.2\% \\
Other languages & 7 & 1.4\% \\
\hline
Total & 509 & 100.0\% \\
\hline
\end{tabular}}
\vspace*{2pt}
\tabnote{\hangindent.5em\ \ (Response rate: 509/526, up to 3 selections allowed).}
\vspace*{2pt}
\tabnote{\hangindent.5em\ \ \textbf{Notes:} ``Other African languages'' includes Setswana, Xhosa, Yoruba, Swahili, and Hausa; ``Other European languages'' includes Finnish, Hungarian, Latvian, Russian, Norwegian, Danish, Swedish, Catalan, Ukrainian, and Lithuanian; ``Other Asian languages'' includes Hindi, Hebrew, Turkish, Arabic, Mongolian, Korean, Japanese, Persian, Bengali, Tagalog, Thai, and Sinhalese; ``Other languages'' includes Portuguese (Brazilian), Slovenian, Croatian, and unspecified others.}
\end{table}

\begin{table}[!t]%%%32
\centering
\tbl{Barriers Accessing ChatGPT or Submitting Feedback\label{tab:barriers}}{
\begin{tabular}{lrr}
\hline
Barriers & Count & Percentage \\
\hline
Barriers & 118 & 22.7\% \\
No barriers & 401 & 77.3\% \\
\hline
Total & 519 & 100.0\% \\
\hline
\end{tabular}}
\vspace*{2pt}
\tabnote{\hangindent.5em\ \ (Response rate: 519/526).}
\end{table}

\begin{table}[!t]%%%33
\centering
\tbl{Frequency of Engaging with AI-related Topics\label{tab:ai-topics-frequency}}{
\begin{tabular}{lrr}
\hline
Frequency & Count & Percentage \\
\hline
Daily & 87 & 16.6\% \\
Weekly & 243 & 46.5\% \\
Monthly & 99 & 18.9\% \\
Rarely & 85 & 16.3\% \\
Never & 9 & 1.7\% \\
\hline
Total & 523 & 100.0\% \\
\hline
\end{tabular}}
\vspace*{2pt}
\tabnote{\hangindent.5em\ \ (Response rate: 523/526).}
\end{table}

\begin{table}[!t]%%%34
\centering
\tbl{Prior Use of AI, ML, or Data Science Tools\label{tab:prior-ai-use}}{
\begin{tabular}{lrr}
\hline
Response & Count & Percentage \\
\hline
Yes & 214 & 41.0\% \\
No & 308 & 59.0\% \\
\hline
Total & 522 & 100.0\% \\
\hline
\end{tabular}}
\vspace*{2pt}
\tabnote{\hangindent.5em\ \ (Response rate: 522/526).}
\end{table}

\begin{table}[!t]%%%35
\centering
\tbl{Completed Education or Training in AI, ML, or Data Science\label{tab:ai-education}}{
\begin{tabular}{lrr}
\hline
Response & Count & Percentage \\
\hline
Yes & 138 & 26.5\% \\
No & 382 & 73.5\% \\
\hline
Total & 520 & 100.0\% \\
\hline
\end{tabular}}
\vspace*{2pt}
\tabnote{\hangindent.5em\ \ (Response rate: 520/526).}
\end{table}

\begin{table}[!t]%%%36
\centering
\tbl{Currently or Previously Working in AI, ML, or Data Science\label{tab:ai-work}}{
\begin{tabular}{lrr}
\hline
Response & Count & Percentage \\
\hline
Yes & 111 & 21.3\% \\
No & 410 & 78.7\% \\
\hline
Total & 521 & 100.0\% \\
\hline
\end{tabular}}
\vspace*{2pt}
\tabnote{\hangindent.5em\ \ (Response rate: 521/526).}
\end{table}

\begin{table}[!t]%%%37
\centering
\tbl{Confidence Submitting Feedback\label{tab:feedback-confidence}}{
\renewcommand{\arraystretch}{1.1}
\begin{tabular}{lrr}
\hline
Confidence level & Count & Percentage \\
\hline
Strongly agree & 114 & 21.8\% \\
Somewhat agree & 296 & 56.5\% \\
Neither agree nor disagree & 79 & 15.1\% \\
Somewhat disagree & 30 & 5.7\% \\
Strongly disagree & 5 & 1.0\% \\
\hline
Total & 524 & 100.0\% \\
\hline
\end{tabular}}
\vspace*{2pt}
\tabnote{\hangindent.5em\ \ (Response rate: 524/526).}
\end{table}

\clearpage

\section{Statistical Relationships and Test Results}

Appendix C presents the statistical analysis of relationships between key variables for this study.

\begin{table}[!h]%%%38
\centering
\tbl{Chi-squared Test Results\label{tab:chi-squared-results}}{
\fontsize{8.5}{9.5}\selectfont
\tabcolsep7.5pt
\renewcommand{\arraystretch}{1.1}
\begin{tabular}{lcccccclr}
\hline
Relationship & $\chi^2$ & $p$ & Adj. $p$ & df & CV & Interp. & $n$ \\
\hline
Thumbs-Up/down and After Regenerating & 50.419 & <.001 & <.001* & 1 & .319 & Mod. & 497 \\
Education/Training in AI and Confidence & 12.773 & .002 & .023* & 2 & .160 & Weak & 497 \\
Working in AI and Barriers & 9.821 & .002 & .023* & 1 & .143 & Weak & 477 \\
Region and Open-text feedback & 13.990 & .003 & .023* & 3 & .192 & Mod. & 378 \\
Prior AI use and Open-text feedback & 8.926 & .003 & .023* & 1 & .154 & Weak & 378 \\
Thumbs-Up/down and ChatGPT frequency & 13.689 & .008 & .045* & 4 & .166 & Weak & 497 \\
AI topics frequency and Confidence & 20.726 & .008 & .045* & 8 & .144 & Weak & 497 \\
Age and Confidence & 17.068 & .009 & .045* & 6 & .131 & Weak & 497 \\
Education/Training in AI and Open-text & 6.132 & .013 & .059 & 1 & .127 & Weak & 378 \\
Gender and After Regenerating & 10.427 & .015 & .061 & 3 & .145 & Weak & 497 \\
Age and Barriers & 9.372 & .025 & .078 & 3 & .140 & Weak & 477 \\
Open-text and ChatGPT frequency & 10.956 & .027 & .078 & 4 & .170 & Mod. & 378 \\
AI topics frequency and Open-text & 10.937 & .027 & .078 & 4 & .170 & Mod. & 378 \\
Region and After Regenerating & 9.151 & .027 & .078 & 3 & .136 & Weak & 497 \\
Working in AI and Open-text feedback & 4.660 & .031 & .080 & 1 & .111 & Weak & 378 \\
Age and After Regenerating & 8.678 & .034 & .080 & 3 & .132 & Weak & 497 \\
Prior AI use and Barriers & 4.512 & .034 & .080 & 1 & .097 & Negl. & 477 \\
Education/Training in AI and Barriers & 4.010 & .045 & .101 & 1 & .092 & Negl. & 477 \\
Gender and Confidence & 12.513 & .051 & .103 & 6 & .112 & Weak & 497 \\
Prior AI use and Confidence & 5.953 & .051 & .103 & 2 & .109 & Weak & 497 \\
Working in AI and Confidence & 5.632 & .060 & .114 & 2 & .106 & Weak & 497 \\
Open-text and After Regenerating & 3.424 & .064 & .117 & 1 & .095 & Negl. & 378 \\
AI topics frequency and Thumbs-Up/down & 7.615 & .107 & .186 & 4 & .124 & Weak & 497 \\
Working in AI and After Regenerating & 2.375 & .123 & .205 & 1 & .069 & Negl. & 497 \\
Regenerating and ChatGPT frequency & 5.004 & .287 & .459 & 4 & .100 & Weak & 497 \\
Age and Open-text feedback & 3.342 & .342 & .526 & 3 & .094 & Weak & 378 \\
Gender and Thumbs-Up/down & 3.160 & .368 & .545 & 3 & .080 & Weak & 497 \\
AI topics frequency and Barriers & 3.962 & .411 & .554 & 4 & .091 & Weak & 477 \\
AI topics frequency and Regenerating & 3.932 & .415 & .554 & 4 & .089 & Weak & 497 \\
Prior AI use and Thumbs-Up/down & 0.666 & .415 & .554 & 1 & .037 & Negl. & 497 \\
Gender and Barriers & 2.482 & .478 & .607 & 3 & .072 & Weak & 477 \\
Education/Training in AI and Thumbs & 0.468 & .494 & .607 & 1 & .031 & Negl. & 497 \\
Working in AI and Thumbs-Up/down & 0.453 & .501 & .607 & 1 & .030 & Negl. & 497 \\
\hline
\multicolumn{8}{r}{\hspace*{5pt}(Continued)}\\
\end{tabular}}
\end{table}

\setcounter{table}{37}
\begin{table}[!h]%%%38
\centering
\tbl{Continued\label{tab:chi-squared-results}}{
\fontsize{8.5}{9.5}\selectfont
\tabcolsep7.5pt
\renewcommand{\arraystretch}{1.1}
\begin{tabular}{lcccccclr}
\hline
Relationship & $\chi^2$ & $p$ & Adj. $p$ & df & CV & Interp. & $n$ \\
\hline
Gender and Open-text feedback & 1.801 & .615 & .723 & 3 & .069 & Weak & 378 \\
Age and Thumbs-Up/down & 1.713 & .634 & .725 & 3 & .059 & Negl. & 497 \\
Prior AI use and After Regenerating & 0.140 & .708 & .773 & 1 & .017 & Negl. & 497 \\
Region and Barriers & 1.361 & .715 & .773 & 3 & .053 & Negl. & 477 \\
Education/Training in AI and Regenerating & 0.065 & .799 & .841 & 1 & .011 & Negl. & 497 \\
Region and Thumbs-Up/down & 0.708 & .871 & .894 & 3 & .038 & Negl. & 497 \\
Region and Confidence & 1.548 & .956 & .956 & 6 & .039 & Negl. & 497 \\
\hline
\hline
\end{tabular}}
%\small
\vspace*{2pt}
\tabnote{\hangindent.5em\ \
\textbf{Notes:}}
\begin{itemize}
  \item[\mdash] Adj. $p$ = $p$-value after Benjamini-Hochberg procedure; df = Degrees of Freedom; CV = Cramér's V; Interp. = Interpretation; Mod. = Moderate; Negl. = Negligible, $n$ = Sample size after removing missing values.
  \item[\mdash] Table organised in order of adj. $p$-value.
  \item[\mdash] * Indicates statistically significant $p$-value after adjustment.
  \item[\mdash] $\chi^2$ test was not performed between Thumbs-Up/down and Open-text feedback. Open-text feedback was only available to users who provided Thumbs-Up/down feedback, making these variables dependent.
  \item[\mdash] Table \ref{tab:cramers-v-interpretation} provides a guide for interpreting Cramér's V values, in accordance with \cite{Cohen2013-ae}. Thresholds for df $\geq$ 3 are approximate guidelines.
\end{itemize}
\end{table}

\begin{table}[!h]%%%39
\vspace*{6pt}
\centering
\tbl{Interpretation of Cramér's V\label{tab:cramers-v-interpretation}}{
\renewcommand{\arraystretch}{1.1}
\begin{tabular}{lrrr}
\hline
Association & df = 1 & df = 2 & df $\geq$ 3 \\
\hline
Weak & 0.10 - <0.30 & 0.07 - <0.21 & 0.06 - <0.17 \\
Moderate & 0.30 - <0.50 & 0.21 - <0.35 & 0.17 - <0.29 \\
Strong & $\geq$0.50 & $\geq$0.35 & $\geq$0.30 \\
\hline
\end{tabular}}\vspace*{8pt}
\end{table}

\clearpage

\end{document}